\documentclass[nonacm,acmsmall,a4paper,screen]{acmart}

\usepackage{booktabs,array}
\usepackage[T1]{fontenc}
\usepackage{microtype}
\usepackage{algorithm}
\usepackage{tikz}
\usetikzlibrary{decorations.pathreplacing,patterns}
\usepackage{cleveref}

\newtheorem{remark}{Remark}

\newcommand{\lgv}{\operatorname{lg}}
\newcommand{\bits}{\{0,1\}}
\newcommand{\eps}{\varepsilon}
\newcommand{\set}[1]{\{#1\}}
\newcommand{\ceil}[1]{\left\lceil #1\right\rceil}
\newcommand{\floor}[1]{\left\lfloor #1\right\rfloor}

\newcommand{\polylog}{\operatorname{polylog}}

\title[Beating Rytter's Bound]{Beating Rytter's Bound:\\
All-Pairs CFL Reachability in $O(n^3(\lgv\lgv n)^2/(\lgv n)^2)$ Time}
\author{Rupak Majumdar}
\affiliation{
\institution{Max Planck Institute for Software Systems}
\city{Kaiserslautern}
\country{Germany}
}
\email{rupak@mpi-sws.org}

\begin{abstract}
Context-free reachability and 2NPDA recognition are fundamental problems in program analysis and language theory.
Despite years of effort, the best algorithms for these problems run
in time $O(n^3/\lgv n)$ on a word RAM model; 
they are based on an observation of Rytter in 1985 for 2NPDA which was adapted for CFL reachability by Chaudhuri in 2008.
No better algorithm has been designed since.

We give the first improvement to Rytter's algorithm.
For both  all-pairs context-free reachability and 2NPDA recognition, we 
show a deterministic algorithm running in
$O(n^3(\lgv\lgv n)^2/(\lgv n)^2)$ time and
$O(n^{2+\eps})$ words of space for every fixed $0<\eps<1$ on a word RAM.
Our improvement is based on a new data structure that reports set differences using batched 
maintenance of exact intersection counts, 
allowing saturation to skip regions that contain no new reachability information.
\end{abstract}

\begin{document}
\maketitle

\section{Introduction}
\label{sec:intro}

Many problems in inter-procedural program analysis, analysis of recursive state machines, and database theory reduce to context-free reachability (CFL reachability)
\cite{DolevEK82,Yannakakis90,sharir-pnueli,MelskiReps,rhs,BEM97,AlurBEGRY05,HeintzeMcAllester}.
For a fixed context-free grammar $\mathcal G$ over an alphabet $\Sigma$, and an input directed graph $G=(V, E)$ whose edges are labeled with letters from $\Sigma$,
the \emph{all-pairs context-free reachability problem} asks to output a relation $R_A \subseteq V\times V$ for each nonterminal $A$ of $\mathcal G$ such that $R_A(u, v)$ iff there is a path
from $u$ to $v$ labeled by a word derivable from the nonterminal $A$.
CFL reachability has a classical algorithm based on saturation \cite{Yannakakis90,rhs}: the  algorithm records facts $A(u,v)$ saying that a walk from $u$ to
$v$ has a label derivable from nonterminal $A$. 
A production $A\to BC$ combines $B(u,z)$ with $C(z,v)$ and adds $A(u, v)$ to the set of facts.
The algorithm terminates when no new facts can be derived.
There are quadratically many possible facts for a fixed grammar but finding their consequences by scanning
all third vertices takes cubic time; thus, naive saturation gives a cubic algorithm.

A related problem in language theory is \emph{2NPDA recognition}.
Given a fixed 2NPDA and an input word $w$ of length $\ell$, it asks whether the (fixed) 2NPDA accepts $w$ \cite{GrayHI67,HarrisonI68}.
An algorithm for 2NPDA recognition saturates paths in the automaton, its run time is again cubic.
The equivalent run times are not an accident:
2NPDA recognition and CFL reachability are linear-time inter-reducible \cite{HeintzeMcAllester,cms}.
In fact, these are two representative problems from a class of many
problems in program analysis 
for which the natural algorithms run in cubic time
(the ``cubic bottleneck'').
These problems include slicing, dataflow, heap-shape, points-to analysis, and set constraints 
\cite{reps-survey,cms,BalasubramanianCM25,kodumal-aiken,Neal,HeintzeMcAllester}.

A natural question is whether the cubic bottleneck can be improved.
In 1985, Rytter showed that a logarithmic improvement is possible \cite{rytter}.
He obtained an $O(\ell^3/\lgv\ell)$-time algorithm for 2NPDA recognition, which implements the saturation procedure on top of packed-set data structures. 
Chaudhuri applied the corresponding speedup to CFL and recursive-state-machine reachability
to obtain an algorithm for CFL reachability that runs in time $O(n^3/\lgv n)$ for a graph with $n$ vertices~\cite{chaudhuri}.
No better algorithm is known and breaking Rytter's bound in the general case is a longstanding open problem (see, e.g., \cite{pavlogiannis-survey}).
There are barriers from fine-grained complexity to finding truly subcubic algorithms 
(algorithms running in time $O(n^{3-\delta})$ for some constant $\delta>0$)
for this class, 
\cite{nfa,cms}.
However, obtaining a polylogarithmic improvement is consistent with fine-grained barriers, which only obstruct truly sub-cubic algorithms.
But, no such improvement was found in 40 years.

In this paper, we show a new algorithm for all-pairs CFL reachability 
that improves Rytter's bound by a factor $\Theta(\lgv n/(\lgv\lgv n)^2)$.

\begin{theorem}[All-pairs CFL reachability]\label{thm:main-intro}
For every fixed context-free grammar and fixed $0<\eps<1$, all-pairs
CFL reachability on an $n$-vertex directed labeled graph can be computed
in
\[
 O\left(\frac{n^3(\lgv\lgv n)^2}{\eps(\lgv n)^2}\right)
\]
deterministic time and $O(n^{2+\eps})$ words of space on a word RAM with
$\Theta(\lgv n)$-bit words. 
\end{theorem}

Our algorithm explicitly returns the Boolean
relation $R_A$ for every nonterminal $A$, including the start symbol.
As an immediate corollary, we can also decide CFL reachability between a designated start and an end state as well as 2NPDA recognition.
(Note that the space requirement for our algorithm is $O(n^{2+\eps})$ for any fixed $0 < \eps < 1$; the standard algorithm requires quadratic space.)

\smallskip\noindent\textbf{Main ideas}\label{sec:ideas}
The saturation algorithm for CFL reachability repeatedly performs the following step: 
When it learns a new fact $B(u,v)$ and the grammar has a rule $A\to BC$, it
finds all vertices $x$ with $C(v,x)$ already known but $A(u,x)$ not
yet known.
Symmetrically, when it learns a new fact $C(v, x)$,  it finds all $u$ such that $B(u, v)$ is known but $A(u, x)$ not known.

Each of these steps requires querying a set difference $X\setminus Y$ of two sets of vertices. 
With bit vector representations of sets, one difference costs about $n/\lgv n$ word
operations, and there are about $n^2$ such questions. 
This gives the $O(n^3/\lgv n)$ bound of Rytter and Chaudhuri. 
Across all difference queries, only $O(n^2)$ elements are reported, because each report produces a previously unknown fact. 
Nevertheless, the packed-set implementation performs a linear scan for every query, even when its answer is empty.
Any improvement to Rytter's bound has to bound this wasted work.

Our main technical contribution is a data structure for reporting set differences
between two families of sets that grow independently during saturation.
It extends the sparse-subset tabulation principle \cite{blelloch} to exact integer counts.
The data structure combines the following three ideas.

\paragraph{Idea 1: count before you scan.}
We divide the $n$ vertices of the graph into about $\lgv n$ groups. 
For each choice of sets $X$ and $Y$ in the saturation algorithm and for each
group, we keep track of how many elements of $X$ in that group also lie in $Y$. 
If this count equals the number of elements of $X$ in the group, the group holds
nothing new and is skipped in the saturation procedure. 
Only groups that are certain to contain a new fact are scanned. 
Each scan therefore pays for itself with a new
fact, and there are only $O(n^2)$ new facts in total.

\paragraph{Idea 2: update the counts lazily.}
Keeping the counts exact after every insertion is expensive. 
A new element of a set changes its counts against all $n$ sets of the other side. 
Our idea is to make new elements first wait in a short pending list.
Subsequent queries inspect this list directly. 
Only after about $n/(\lgv n)^2$ elements
have accumulated in the pending list are the counts brought up to date, in one batch.

\paragraph{Idea 3: precompute small pieces of a batch.}
A batch update must add the contributions of many elements to the
counts of all $n$ opposite sets. 
We precompute these contributions for every small set of coordinates, of 
size up to about $\lgv n/\lgv\lgv n$, inside a block of coordinates. 
A batch is split into such small pieces, and each piece costs one table lookup and one
word-parallel vector addition instead of one addition per element.
Because only small sets are tabulated, the blocks can be much longer
than $\lgv n$ while the tables stay polynomial in size.
This ensures that we maintain small space while answering queries.

One subtlety is that, as saturation progresses, both $X$ and $Y$ grow. 
An element that belongs to both sets must be counted exactly once, 
when the later of its two batches is processed.
For each set, we maintain a snapshot containing the elements whose pending batches have already been processed.
The key is maintaining snapshots on both sides of a query. 
We split the stored counts between two arrays, 
one supporting word-parallel row updates and the other
supporting word-parallel column updates. 
Short pending lists correct the snapshot counts exactly, allowing the algorithm to identify regions
that contain a new fact before scanning them. 
Every such scan can therefore be charged to a newly derived fact. 

\paragraph{Outcome.}
Together, these ideas save a factor of 
$\lgv n/(\lgv\lgv n)^2$ over the packed-set algorithm. 
The two $\lgv\lgv n$ factors come from the size of the small pieces and from
the bits needed to store their counts. The result is the bound of
Theorem~\ref{thm:main-intro}. 

Our contribution is a purely theoretical analysis.
We do not claim that  our algorithm will outperform existing, optimized implementations of CFL reachability.
However, we consider breaking Rytter's bound an important breakthrough for the problem
which may open the door to further improvements.

\section{Related Work}\label{sec:related}

\smallskip\noindent\textbf{Practical CFL-reachability algorithms}
A lot of research has gone into making CFL reachability faster in practice, by 
exploiting either client needs or the program structure, or by providing
engineering improvements that scale in practice but do not provide a better worst-case
complexity.

Demand-driven analyses avoid computing facts unrelated to a client's
queries. Horwitz, Reps, and Sagiv~\cite{hrs-demand} formulate
interprocedural dataflow analysis around requests for individual facts
at individual program points. Zheng and Rugina~\cite{zheng-rugina}
answer alias queries by exploring a CFL-reachability graph until enough
evidence is found, without first constructing and intersecting complete
points-to sets. These approaches exploit selective queries; however,
they do not break the cubic bottleneck.

Systems work addresses the cost of storing and processing large analysis
graphs. Graspan~\cite{graspan} uses disk-based graph processing to scale
interprocedural analyses beyond main memory. Chianina~\cite{chianina}
adds an evolving-graph formulation for flow sensitivity and disk support
for the memory demands of context sensitivity. These systems separate
analysis logic from the mechanisms used to execute it efficiently.

Other methods reduce redundant derivations. The partially ordered
solver of Lei et al.~\cite{pocr} maintains spanning-tree representations
of transitive relations to avoid repeatedly deriving the same
consequences. Graph folding~\cite{graph-folding} uses a recursive
state-machine representation of the grammar to guide reductions of the
input graph. Skewed tabulation~\cite{skewed} changes how recursive
productions are evaluated, through grammar transformations or a
specialized solver, to avoid intermediate summary edges that do not
contribute to the requested relation. Our structure addresses a
different source of work within saturation: testing set differences
that contain few or no new facts. Understanding how its batching costs
interact with these evaluation strategies is a question for future
implementations.

\smallskip\noindent\textbf{Faster algorithms under restrictions}\label{sec:restricted}
Restrictions on the language, graph, or machine can change the
complexity substantially~\cite{pavlogiannis-survey}. For Dyck-1,
stack height determines stack content, since there is only one
parenthesis type. Mathiasen and Pavlogiannis~\cite{mathiasen} give an
$O(n^\omega\lgv^2 n)$ all-pairs algorithm; related algebraic bounds
were obtained by Bradford~\cite{Bradford2018}. 
General CFL reachability reduces to Dyck-2 reachability, for which the above algorithms do not work.
Grammar-specific classifications also distinguish tractable languages from harder
ones~\cite{koutris-deep}.

In bidirected Dyck graphs, each opening edge has a reverse edge with
the matching closing label. Reachability then forms an equivalence
relation, allowing algorithms to maintain classes of vertices.
Following earlier work~\cite{zhang-pldi13,zhang-oopsla14}, Chatterjee,
Choudhary, and Pavlogiannis~\cite{ChatterjeeCP18} compute this compact
representation in $O(m+n\alpha(n))$ time, where $m$ counts edges and
$\alpha$ is the inverse Ackermann function. Explicitly listing all
reachable pairs can still require quadratic output. Treewidth offers
another restriction: Chatterjee et al.~\cite{cipg15} exploit constant
treewidth of recursive state-machine components to preprocess
interprocedural queries. This differs from assuming constant treewidth
of an arbitrary Dyck graph, which retains combinatorial Boolean
matrix-multiplication hardness~\cite{ChatterjeeCP18}.

Restrictions on computations give further improvements. 
On loop-free 2NPDAs~\cite{rytter} and bounded-stack recursive
state machines~\cite{chaudhuri}, one can get an $O(n^3/(\lgv n)^2)$ algorithms.
For fixed \emph{deterministic} two-way pushdown automata, Cook showed a linear time
recognition algorithm~\cite{cook,gluck}, famously used in Knuth-Morris-Pratt's algorithm
for string matching. 
However, his procedure does not establish the same bound for 2NPDA recognition.

\smallskip\noindent\textbf{Parsing and matrix-vector algorithms}\label{sec:related-bmm}
A different direction of research relates CFL problems with matrix multiplication.
Valiant's parser~\cite{valiant} uses Boolean matrix multiplication to
combine facts about substrings. Substring intervals give the recursive
computation an order that arbitrary cyclic graphs do not provide.
For a static Boolean product, Four Russians tabulation~\cite{ADKF70}
precomputes combinations of small groups of coordinates; together with
word packing, it gives $O(n^3/(\lgv n)^2)$ time. Subsequent combinatorial
algorithms obtain larger savings~\cite{bansal-williams,chan-four-russians,yu-bmm},
including a recent $O(n^3/2^{\Omega{(\lgv n)^{1/7}}})$ algorithm~\cite{afklm-bmm}.

Preprocessing is also useful when the second operand arrives later.
Williams~\cite{williams-mv} preprocesses a fixed matrix over a fixed
finite semiring in $O(n^{2+\eps})$ time to support vector products in
$O(n^2/(\eps\lgv n)^2)$ time. In online Boolean matrix--vector
multiplication, each answer must be returned before the next vector
arrives. Larsen and Williams~\cite{larsen-williams} process $n$ such
vectors in randomized time $n^3/2^{\Omega(\sqrt{\lgv n})}$.
Dudek and Gawrychowski~\cite{online-parsing} transfer online-product
improvements to online context-free string recognition. Dynamic
variants also handle changing matrices and adaptive queries~\cite{liu}.
These results make matrix products promising tools for saturation,
but a transfer must match the operations and their frequency. The
discussion at the end of Section~\ref{sec:cfl} explains the additional
requirements of correcting snapshots and discovering missing facts.

\smallskip\noindent\textbf{Sparse tabulation and set data structures}\label{sec:related-ds}
The closest tabulation precedent is Blelloch, Vassilevska, and
Williams~\cite{blelloch}. They divide matrix columns into blocks and
precompute Boolean products for subsets containing only a few
coordinates. Splitting a sparse query into these subsets reduces the
number of combinations needed; the tables also support replacing a
matrix column. Their applications include sparse transitive closure
and other graph problems. We extend this sparse-subset principle to
integer sums. Unlike Boolean answers or a fixed finite semiring,
exact counts have a range growing with the input. Our analysis charges
for their representation width, carries, and table maintenance, and
combines these tables with independently growing snapshots on both
sides of a set query.

Dynamic set-intersection structures provide a different tradeoff.
Kopelowitz, Pettie, and Porat~\cite[Theorem~1]{kpp-dynamic} give a
randomized structure with constant expected update time. On a
logarithmic-word RAM its reporting time is
$O(d\lgv\lgv n/\lgv n+z)$ in expectation, where $d$ bounds set size
and $z$ counts outputs. Their applications include triangle enumeration.
A difference can be written as an intersection
with a complement; insertions into the subtracted set then become
deletions from its complement. Our insertion-only
structure spends more time on updates and more space to obtain
deterministic exact counts and faster empty-difference queries for
dense sets. Static structures additionally support set disjointness
and union--intersection expressions~\cite{cohen-porat,bille-pagh-pagh}.
Batching has precedents in static-to-dynamic transformations~\cite{bentley-saxe};
semi-naive evaluation likewise organizes deductions around newly added
facts~\cite{bancilhon-ramakrishnan}.

\smallskip\noindent\textbf{Conditional lower bounds}
CFL reachability has been studied through the lens of fine-grained complexity.
For string parsing, Lee's BMM reduction uses an input-dependent
grammar~\cite{lee-jacm}, whereas Abboud, Backurs, and Vassilevska
Williams obtain $k$-Clique barriers for a fixed grammar~\cite{abv-parser}. 
Chatterjee, Choudhary, and Pavlogiannis show a reduction from combinatorial BMM to CFL 
reachability~\cite{ChatterjeeCP18}. 
Under the NFA Acceptance Hypothesis~\cite{nfa}, there is no truly subcubic $O(n^{3-\delta})$ 
algorithm for CFL reachability.
These conditional bounds do not preclude logarithmic improvements.
Chistikov, Majumdar, and
Schepper~\cite{cms} show subcubic verification of certificates for both
reachability and non-reachability.
Such certificates establish a distinction
between checking a supplied answer and constructing the reachability
relation.

\section{Preliminaries: All-Pairs CFL Reachability}\label{sec:prelim}

We write $[n]=\set{0,\ldots,n-1}$. A \emph{labeled graph} 
$G=(V,E)$, 
over a (fixed) alphabet $\Sigma$
consists of a (finite) set $V$ of vertices, with $|V| = n$, and labeled edges
$E\subseteq V\times\Sigma\times V$. 
A \emph{walk} is a sequence of vertices
\[
v_0 \xrightarrow{w_0} v_1 \xrightarrow{w_1} \ldots \xrightarrow{w_k} v_{k+1} 
\]
such that $(v_i, w_i, v_{i+1})\in E$ for each $i\geq 0$.
Thus, a walk may repeat vertices and edges. 
The label of a walk is the word formed by the concatenation $w_0w_1\ldots$ of its edge labels; 
the empty walk has label $\eps$.

\begin{definition}[All-pairs CFL reachability]
\label{def:allpairs}
Let $\mathcal G=(N,\Sigma,\mathcal P,S)$ be a context-free grammar (CFG),
where $N$ is its set of nonterminals, $\mathcal P$ its productions,
and $S$ its start symbol. For each $A\in N$, let $L_A$ be the
language derivable from $A$, and define the relation
\[
 R_A=\set{(u,v)\in V\times V:
   \text{some walk from $u$ to $v$ has a label in }L_A}.
\]
The \emph{all-pairs CFL reachability problem} takes as input a graph $G$ and
a CFG $\mathcal G$ and outputs, for every nonterminal $A\in N$, the
Boolean matrix of $R_A$: for every ordered pair $(u,v)\in V\times V$,
its entry is one if and only if $(u,v)\in R_A$. 
The output includes diagonal pairs $(v, v)$ and
accounts for empty walks when $\eps\in L_A$. 
We consider the grammar to be fixed in our principal bound; 
the graph is the varying input.
For a fixed grammar the output matrices have $O(n^2)$ Boolean entries in total. 
\qed
\end{definition}

The \emph{(designated-pair) CFL reachability problem} additionally
takes two vertices $s,t\in V$ and outputs the single bit indicating whether
$(s,t)\in R_S$.
Clearly, the complexity of CFL reachability is at most that of all-pairs CFL reachability
(but could be asymptotically better).
Our algorithm solves the all-pairs problem, returning $R_A$ for every
$A\in N$, and also solves the designated-pair problem.

We assume that the grammar $\mathcal G$ is in Chomsky normal form,
that is, its productions are of the forms $A\to BC$, $A\to B$, $A\to a$, and
$A\to\eps$. 
In statements with grammar dependence, $|\mathcal G|$ refers to this normalized 
representation (which is polynomial in the size of the original grammar),
and we write $P$ for its number of binary productions. 
A \emph{fact} $A(u,v)$ is an entry in a nonterminal relation. 
If $A \to a$, then $A(u, v)$ holds for each $(u, a, v) \in E$,
and if $A \to \eps$, then $A(v,v)$ holds for all $v\in V$.

The Dyck language with $k$ types consists of properly matched words in
openings $(_1,\ldots,(_k$ and closings $)_1,\ldots,)_k$. It is generated by
$S\to\eps\mid SS\mid (_iS )_i$. 
For fixed $k$, its normalization has constant size. 
It is known that Dyck-1 reachability can be solved in $\tilde{O}(n^\omega)$ time \cite{mathiasen},
and any all-pairs reachability problem for a fixed grammar can be reduced to Dyck-2
reachability in linear time in the bitsize of the input \cite{cms}.

\paragraph{Computational model.}
We use a word RAM with Boolean operations, shifts, integer arithmetic,
and random access to words. Each operation on one word costs unit time,
and space is measured in words.
The word size is $w=\Theta(\lgv n)$, with a fixed constant large enough
to address all the memory we use. This memory is polynomial in $n$.

If the grammar is part of the input, memory also holds one matrix per
nonterminal, so an address needs $\Theta(\lgv(n+|\mathcal G|))$ bits.
In that case we take $w=\Theta(\lgv(n+|\mathcal G|))$.
Our data structures never pack more than $\Theta(\lgv n)$ bits of data
into one word, even when $w$ is larger. Extra bits in a word are used
only for addresses and do not provide additional parallelism. All bounds
are therefore stated in terms of $\lgv n$ and not in terms of~$w$.
For a fixed grammar, $\lgv(n+|\mathcal G|)=\Theta(\lgv n)$, so the two
choices of word size agree.

\paragraph{Packed sets.}
A \emph{packed set} represents a subset $S\subseteq[n]$ by its
characteristic bit vector, stored in $\ceil{n/w}$ consecutive words.
Element $i$ is bit $i\bmod w$ of word $\floor{i/w}$.
Rows and columns of Boolean matrices are stored as packed sets.

\begin{lemma}[Packed sets]\label{lem:packed}
After preprocessing that takes $O(\sqrt n\lgv n)$ time and space,
packed sets over $[n]$ support the following operations.
\begin{enumerate}
\item Membership test, insertion, and deletion of one element in $O(1)$ time.
\item Union, intersection, and difference $S\setminus T$ of two packed
 sets, producing a packed set, in $O(\ceil{n/w})$ time.
\item Enumeration of the elements of $S$ in $O(\ceil{n/w}+|S|)$ time.
\end{enumerate}
In particular, the elements of $S\setminus T$ can be listed in
$O(\ceil{n/w}+|S\setminus T|)$ time.
\end{lemma}
\begin{proof}
For~(1), we compute $\floor{i/w}$ and $i\bmod w$, and test, set, or clear
the bit using a one-bit mask built by a shift. For~(2), we apply the
corresponding bitwise operation word by word. The difference is
$S$ \textsc{and} (\textsc{not} $T$).
For~(3), we scan the words and skip each zero word in $O(1)$ time.
For a nonzero word we need the positions of its one bits. So, we split the
word into chunks of at most $\floor{(\lgv n)/2}$ bits. Since
$w=\Theta(\lgv n)$, there are $O(1)$ chunks. During preprocessing,
we build a table that lists, for every chunk value
$x<2^{\floor{(\lgv n)/2}}\le\sqrt n$, the positions of the one bits
of~$x$. The table has at most $\sqrt n$ entries, each a list of at most
$(\lgv n)/2$ positions, so it takes $O(\sqrt n\lgv n)$ time and space.
We extract each chunk by a shift and a mask, look up its list, and add the
chunk's offset. A nonzero word thus costs $O(1)$ plus the number of
its elements, which gives the bound.
(A word RAM with a constant-time instruction for the lowest set bit
would not need the table. However, we do not assume such an instruction.)
\end{proof}

We use the following conventions when we talk about data structures later.
All stated time bounds include the time to build tables such as the
one in Lemma~\ref{lem:packed}. Bounds are asymptotic. Inputs with $n$
below a fixed constant are handled by brute force in constant time,
so we may assume that $n$ is large enough whenever convenient.

\paragraph{The saturation algorithm.}
The relations $R_A$ computed by all-pairs CFL reachability 
are the least solution of a system of equations.
To see this, we first represent each relation as a Boolean $n\times n$ matrix. 
Let $E_A$ contain the pairs $(u,v)$ with $(u,a,v)\in E$ for some rule $A\to a$,
together with the diagonal $(u, u)$ when $A\to\eps$ is a rule. 
Then the matrices $(R_A)_{A\in N}$ form the least simultaneous solution of
\begin{equation}\label{eq:fixpoint}
 R_A=E_A\vee\bigvee_{A\to B}R_B
           \vee\bigvee_{A\to BC}R_B R_C,
\end{equation}
where multiplication is Boolean and $\vee$ is entrywise \textsc{or}.

The classical algorithm for all-pairs CFL reachability computes this
least solution by saturation~\cite{rhs}. It starts with the initial facts in $E_A$
and repeatedly derives new facts until no rule applies. 
Pending facts are kept in a queue $Q$, called the \emph{worklist}. 
During saturation, $R_A$ denotes the set of facts $A(u,v)$ found so far. The algorithm has no
designated source or target and treats all ordered vertex pairs alike.

When a fact $B(u,v)$ is processed and $A\to BC$ is a rule, the new
conclusions are the facts $A(u,x)$ with $C(v,x)$ already known and
$A(u,x)$ not yet known. In matrix terms, the candidates $x$ form the
\emph{row difference}
\begin{equation}\label{eq:diff-trigger}
 R_C[v,:]\setminus R_A[u,:].
\end{equation}
Symmetrically, when a fact $C(v,x)$ is processed, the new conclusions
$A(u,x)$ are given by the \emph{column difference}
$R_B[:,v]\setminus R_A[:,x]$.

\begin{algorithm}[t]
\begin{quote}\small
\textsc{Insert}$(Z,a,b)$:\\
\hspace*{1em}If $(a,b)\in R_Z$, return.\\
\hspace*{1em}Add $(a,b)$ to $R_Z$ and append $(Z,a,b)$ to $Q$.\\[3pt]
\textsc{Saturate}:\\
\hspace*{1em}For each rule $A\to a$ and each edge $(u,a,v)\in E$, call
 \textsc{Insert}$(A,u,v)$.\\
\hspace*{1em}For each rule $A\to\eps$ and each $v\in V$, call
 \textsc{Insert}$(A,v,v)$.\\
\hspace*{1em}While $Q$ is nonempty, remove its first element $(Z,a,b)$.\\
\hspace*{2em}For each rule $A\to Z$, call \textsc{Insert}$(A,a,b)$.\\
\hspace*{2em}For each rule $A\to ZC$, compute $D=R_C[b,:]\setminus R_A[a,:]$;\\
\hspace*{3em}for every $x\in D$, call \textsc{Insert}$(A,a,x)$.\\
\hspace*{2em}For each rule $A\to BZ$, compute $D=R_B[:,a]\setminus R_A[:,b]$;\\
\hspace*{3em}for every $u\in D$, call \textsc{Insert}$(A,u,b)$.\\
\hspace*{1em}Return $(R_A)_{A\in N}$.
\end{quote}
\caption{The algorithm \textsc{Saturate}.
Each difference $D$ is computed completely before its elements are
inserted. This matters for rules such as $A\to AC$, where the
insertions change the set being subtracted.
\label{algo:saturate}
}
\end{algorithm}

\Cref{algo:saturate} shows the saturation algorithm. We summarize its properties below (see, e.g., \cite{rhs}).

\begin{lemma}[Saturation]\label{lem:saturate}
\textsc{Saturate} terminates. For every nonterminal $A$ and every
ordered pair $(u,v)$, the returned matrix satisfies $R_A[u,v]=1$ if
and only if some walk from $u$ to $v$ has a label in $L_A$.
\end{lemma}
\begin{proof}
\textsc{Insert} adds each fact at most once, and there are at most
$|N|n^2$ facts. So $Q$ receives finitely many elements and the loop
terminates.
Soundness: every inserted fact has a witness walk. The base cases are witnessed by single
edges or empty walks. 
A unit rule $A \rightarrow B$ maintains the witness. 
A binary rule concatenates the witnesses of its two premises.

Completeness: we show that the returned relations satisfy every rule.
Equation~\eqref{eq:fixpoint} then implies that they contain the least
solution. Seed edges are inserted at the start. Every inserted fact is later
processed, and processing it applies all unit rules to it. Now take a
binary rule $A\to BC$ and returned facts $B(u,v)$ and $C(v,x)$. Consider
whichever of the two is processed later. If it is $B(u,v)$, then
$x\in R_C[v,:]$ at that moment, so either $A(u,x)$ is already known or
$x$ is in the row difference and $A(u,x)$ is inserted. The case of
$C(v,x)$ is symmetric and uses the column difference.
\end{proof}

\paragraph{Cost of saturation.}
Let $F=\sum_{A\in N}|R_A|$ be the number of facts at termination; for
a fixed grammar $F=O(n^2)$. Each fact is processed once. For each
binary rule containing its nonterminal on the right-hand side,
processing a fact issues one difference query. So there are
$O(PF)$ difference queries. Every element reported by a query is
absent from $R_A$ when the query starts and is inserted immediately
afterwards. It is therefore never reported again, by this query or by
any later one. Summed over the whole run, the reported elements
number at most~$F$.
\begin{itemize}
\item \emph{Naive implementation.} Computing a difference by testing
 all $n$ coordinates takes $O(n)$ time. The total is $O(PFn+|\mathcal G|F)$.
 For a fixed grammar this is $O(n^3)$.
\item \emph{Packed sets.} Store every row and every column of every
 $R_A$ as a packed set. An insertion updates one row and one column
 in $O(1)$ time. By Lemma~\ref{lem:packed}, a difference with $z$
 reported elements takes $O(n/w+z)$ time. The total is
 $O(PFn/w+|\mathcal G|F)$, which is $O(n^3/\lgv n)$ for a fixed
 grammar.
\end{itemize}
This packed-set speedup is the technique of Rytter for
two-way pushdown automata~\cite{rytter}. Chaudhuri applied it to CFL
reachability~\cite{chaudhuri}. Its $O(n^3/\lgv n)$ bound is the
baseline that we improve.

\paragraph{The data-structure problem.}
The cost of saturation is dominated by the difference queries. Both
operands only grow during the run: rows and columns of each $R_A$ gain
elements as saturation proceeds and never lose them. 
Abstractly, we need to maintain two
families of sets $X[v]$ and $Y[u]$, $u,v\in[n]$, where all sets are subsets of $[n]$.
We want to allow insertions into these sets 
and answer queries
\[
 \text{report } X[v]\setminus Y[u].
\]
For a row instance of $A\to BC$, the sets are $X[v]=R_C[v,:]$ and
$Y[u]=R_A[u,:]$. Column instances are symmetric. 

With packed sets, a query costs $\Theta(n/\lgv n)$ even when the answer is empty.
There can be $\Theta(n^2)$ such queries, which gives the cubic term.
To beat $O(n^3/\lgv n)$, a query reporting $z$ elements should cost
$o(n/\lgv n)$ plus a term that can be charged to the $z$ reported
elements, while insertions remain cheap in amortized terms.
Next, in Sections~\ref{sec:sparse} and~\ref{sec:sets}, we show how to build such a data structure
(Theorem~\ref{thm:sets}). 
Finally, Section~\ref{sec:cfl} plugs the new data structure into \textsc{Saturate} and work out the complexity bounds.

Equation~\eqref{eq:fixpoint} explains the connection of CFL reachability to matrix multiplication.
However, the $O(n^3/(\lgv n)^2)$ Four Russians
bound for one static Boolean product (Section~\ref{sec:related-bmm})
does not carry over to our setting. First, the unknown matrices occur on both sides.
Second, the relations change and the queries are generated adaptively.
Thus, a bound for one matrix product with fixed operands does not bound either the number of
products or the cost of their incremental changes.

\section{Sparse subset tabulation}\label{sec:sparse}

Sections~\ref{sec:sparse} and~\ref{sec:sets} solve the data-structure
problem stated at the end of Section~\ref{sec:prelim}. Since the problem
does not depend on CFL reachability and is of independent interest,
we state it in a general form.

\paragraph{The set-pair problem.}
The goal of the data structure is to maintain two \emph{families} of sets over the universe $[n]$:
$X[v]$ for $v\in[n]$ and $Y[u]$ for $u\in[n]$. 
Equivalently, we can assume that there are two Boolean $n\times n$ matrices whose rows are the 
sets $X[v]$ and the sets $Y[u]$, respectively. 
Each such row is called a \emph{line}:
$X[v]$ is an $X$ line and $Y[u]$ is a $Y$ line. 
The lines of the other family are called the \emph{opposite} lines; 
for an $X$ line, these are the $n$ lines $Y[u]$, and vice versa. 
Initially every line is empty.
The data structure supports two operations.
\begin{itemize}
\item \emph{Insert} an element $x$ into a line, that is, change one
 matrix entry from $0$ to $1$. Entries never change back.
\item \emph{Report} the difference $X[v]\setminus Y[u]$ for a given pair
 $(v,u)$.
\end{itemize}
In \textsc{Saturate}, a binary rule $A\to BC$ gives a row instance with
$X[v]=R_C[v,:]$ and $Y[u]=R_A[u,:]$, and a column instance that is
symmetric. 
As we saw, with packed sets,
a report costs $\Theta(n/\lgv n)$ even when the difference is empty.
Our goal is to make reports with few outputs cheaper, while keeping
insertions cheap on average.
In the following, we will describe the data structure and use a number of parameters,
$g$, $D$, $s$, etc.\ as we go along.
These parameters will be picked later to derive the asymptotic bounds.

\paragraph{Overview of the solution.}
The data structure combines two ideas.
\begin{itemize}
\item \emph{Groups and counts.} We partition the universe $[n]$ into $g$
 \emph{groups} of consecutive elements. If, for each group $G$, we knew
 the count $|X[v]\cap Y[u]\cap G|$, a report could compare it with
 $|X[v]\cap G|$ and scan only the groups that contain an output.
 The parameter $g$ will be picked later.
\item \emph{Snapshots and batched updates.} Maintaining these counts
 exactly after every insertion would be too expensive: one insertion
 into an $X$ line changes its counts with all $n$ opposite lines.
 We therefore split each line into two parts. The \emph{snapshot}
 $X_0[v]\subseteq X[v]$ consists of the elements whose effect on the
 stored counts has already been accounted for. The \emph{pending list}
 $\Delta_X(v)=X[v]\setminus X_0[v]$ holds the remaining elements, which
 were inserted since then. The same holds for $Y_0[u]$ and $\Delta_Y(u)$.
 Counts are stored only for snapshots. Reports correct them by additionally using the
 (short) pending lists. When a pending list reaches a threshold of $D$
 elements, the line is \emph{promoted}: its pending elements join its
 snapshot, the counts of this line with all $n$ opposite lines are
 updated, and the pending list is emptied.
 As with the parameter $g$, the parameter $D$ will be picked later.
\end{itemize}
A promotion is the expensive step. Suppose $Y[u]$ is promoted with
pending list $R=\Delta_Y(u)$, so $|R|=D$. The update needs, for every
group $G$ and every opposite line $X[v]$, the number
$|X_0[v]\cap R\cap G|$. 
Arrange the $X$ snapshots as the rows of a Boolean matrix, the \emph{snapshot matrix} of the $X$ family. 
Then the required numbers, for all $v$ at once, form the vector obtained by
adding the columns of this matrix indexed by $R\cap G$. 
Adding the $D$ columns one at a time costs at least $D$ operations on $n$-entry
vectors per promotion.

This section develops the tool that does this step faster, the
\emph{sparse subset table}. 
It works for any Boolean matrix whose entries change from $0$ to $1$, 
such as a snapshot matrix, which
changes whenever a line of its family is promoted. 
The table adds the columns indexed by $R\cap G$ using about $|R|/s$ vector additions
instead of $|R|$ (Lemma~\ref{lem:sparse}), where $s$ is a suitably chosen
sparsity parameter to be decided later. 
Section~\ref{sec:sets} then
maintains the counts, implements promotions and reports, proves their
correctness, and fixes all parameters.

\paragraph{The tabulation problem.}
Throughout this section, $M\in\bits^{r\times m}$ is a Boolean matrix
with $r$ rows and $m$ columns, whose columns are called
\emph{coordinates}. 
Initially $M$ is zero, and it changes only by setting entries from $0$
to $1$. Fix a partition of its coordinates into groups, as described
below. A query supplies a set $R$ of coordinates as a sorted list and
asks, for every group $G$ meeting $R$, for
the vector
\[
 \sum_{x\in R\cap G}M[:,x]\in\set{0,\ldots,|R|}^r,
\]
whose $i$th entry is the number of coordinates of $R\cap G$ that lie
in row~$i$. Groups disjoint from $R$ have the zero vector implicitly.
The \emph{tabulation problem} is to maintain a representation of $M$
under these insertions and answer these queries efficiently.

In Section~\ref{sec:sets}, we will set $M$ to be the snapshot
matrix of one family, so $r=n$ and $m=\Theta(n)$ after padding,
and $R$ will be the pending list of a
line of the other family that is being promoted.

\paragraph{Parameters.}
The section uses four parameters. Their values will be fixed later.
Until then, all statements hold for arbitrary values.
\begin{itemize}
\item $L$, the \emph{block length}. Coordinates are split into blocks
 of $L$ consecutive coordinates, and tables are built per block.
\item $s\le L$, the \emph{sparsity}. Tables store only subsets of a
 block with at most $s$ elements.
\item $g$, the number of \emph{groups}. These are the groups of the
 outline. Each group is a union of consecutive blocks.
\item $D$, the \emph{promotion size}. A query set $R$ is a promoted
 pending list, so $|R|=D$. $D$ is used only at the end of this
 section, to choose $L$.
\end{itemize}
For orientation, the final choice will be $D=\Theta(n/(\lgv n)^2)$,
$g=\Theta(\lgv n)$, $s=\Theta(\lgv n/\lgv\lgv n)$, and
$L=\Theta(s(\lgv n)^2)$. Blocks are therefore polylogarithmic in size
and much smaller than groups.

\paragraph{Blocks and groups.}
We add zero columns (``padding'') to $M$ until $m$ is a multiple of $L$. 
We number the columns of $M$ by $0,\ldots,m-1$ and 
then partition the columns into blocks of
$L$ consecutive columns. 
A group is a union of consecutive blocks, so every block lies inside exactly one group.

\paragraph{Sparse subsets.}
For integers $1\le s\le L$, let
\[
 B_s(L)=\sum_{j=0}^s\binom Lj,\qquad N_s(L)=B_s(L)-1.
\]
A block has $B_s(L)$ subsets of size at most $s$, since the $j$th
term counts the subsets of size exactly $j$. The term for $j=0$ counts
only the empty set, so $N_s(L)$ of these subsets are nonempty.
By algebra, we have $B_s(L)\le(eL/s)^s$.
The point of the bound is that $B_s(L)$ is far smaller than the
$2^L$ subsets of a block when $s\ll L$. So we can afford long blocks,
provided each lookup uses only a few coordinates of a block.

\paragraph{The table.}
For every nonempty set $S$ of at most $s$ coordinates within one
block, the \emph{sparse table} stores the vector
\[
 V_S=\sum_{x\in S}M[:,x]\in\set{0,\ldots,s}^r.
\]
Its $i$th entry is the number of elements of $S$ in row $i$ of $M$.

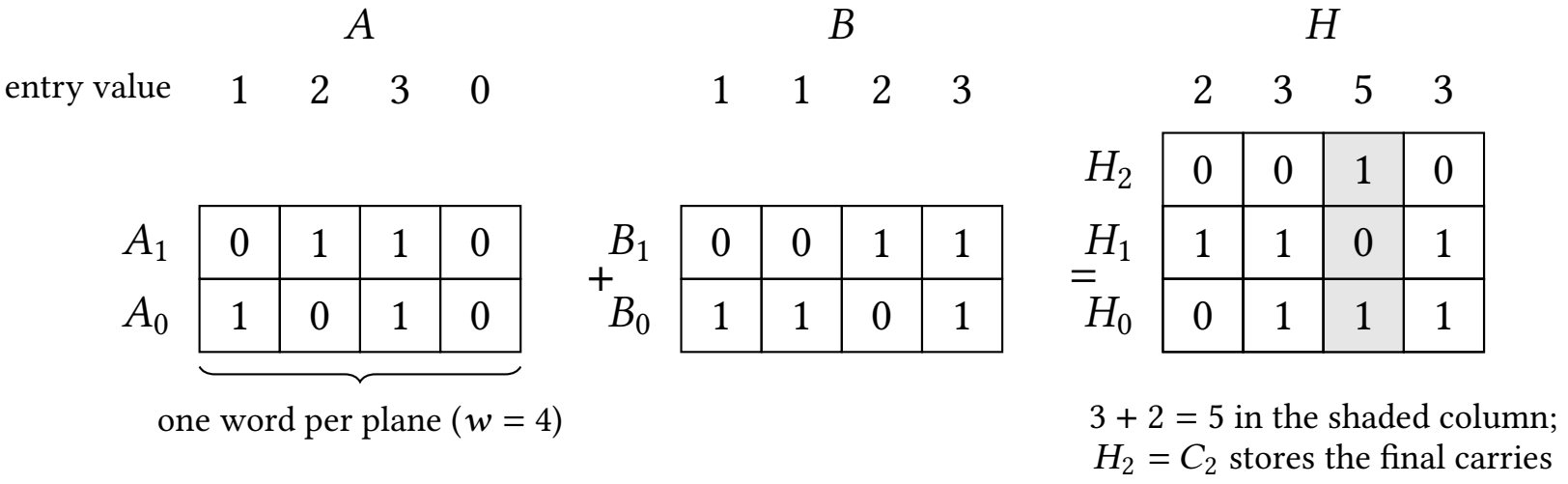
\begin{figure}[t]
\centering
\begin{tikzpicture}[x=0.55cm,y=0.50cm,font=\small]
\foreach \base/\name in {0/A,6/B,12/H} {
  \node at (\base+2,4.5) {$\name$};
  \foreach \j in {0,1} {
    \node[anchor=east] at (\base-0.12,\j+0.5) {$\name_\j$};
    \foreach \i in {0,...,3}
      \draw (\base+\i,\j) rectangle (\base+\i+1,\j+1);
  }
}
\node at (5,1) {$+$};
\node at (11,1) {$=$};
\node[anchor=east,font=\scriptsize] at (-0.12,3.6) {entry value};
\foreach \i/\v/\lo/\hi in {0/1/1/0,1/2/0/1,2/3/1/1,3/0/0/0} {
  \node at (\i+0.5,3.6) {$\v$};
  \node at (\i+0.5,0.5) {$\lo$};
  \node at (\i+0.5,1.5) {$\hi$};
}
\foreach \i/\v/\lo/\hi in {0/1/1/0,1/1/1/0,2/2/0/1,3/3/1/1} {
  \node at (6+\i+0.5,3.6) {$\v$};
  \node at (6+\i+0.5,0.5) {$\lo$};
  \node at (6+\i+0.5,1.5) {$\hi$};
}
\fill[black!10] (14,0) rectangle (15,3);
\foreach \j in {0,...,2}
  \foreach \i in {0,...,3}
    \draw (12+\i,\j) rectangle (12+\i+1,\j+1);
\node[anchor=east] at (11.88,2.5) {$H_2$};
\foreach \i/\v/\lo/\hi/\carry in {0/2/0/1/0,1/3/1/1/0,2/5/1/0/1,3/3/1/1/0} {
  \node at (12+\i+0.5,3.6) {$\v$};
  \node at (12+\i+0.5,0.5) {$\lo$};
  \node at (12+\i+0.5,1.5) {$\hi$};
  \node at (12+\i+0.5,2.5) {$\carry$};
}
\draw[decorate,decoration={brace,mirror,amplitude=3pt}]
  (0,-0.2) -- (4,-0.2) node[midway,below=4pt,font=\scriptsize]
  {one word per plane ($w=4$)};
\node[anchor=north,font=\scriptsize,align=center] at (14,-0.45)
  {$3+2=5$ in the shaded column;\\ $H_2=C_2$ stores the final carries};
\end{tikzpicture}
\caption{Bit-plane addition with $r=w=4$ and two-bit operands
($k=2$). Reading a column from top to bottom gives the binary expansion
of the entry above it. Each horizontal plane fits in one word in this
example, so each Boolean operation in the full-adder equations acts on
all four entries at once. Carries move to the next plane at the same
entry position; they never pass between entries. The sum needs a third
plane. For general $r$, each plane occupies $\ceil{r/w}$ words.}
\Description{Three grids encode A=(1,2,3,0), B=(1,1,2,3), and
their entrywise sum H=(2,3,5,3). Rows are bit planes, with the least
significant plane at the bottom. The third entry of H is shaded and
has bits 101, showing the extra carry plane needed for 3+2=5.}
\label{fig:bit-planes}
\end{figure}

\paragraph{Bit planes.}
Entries of $V_S$ are at most $s$, so they need
$k=\ceil{\lgv(s+1)}$ bits. We store $V_S$ in $k$ \emph{bit planes}:
plane $j$ packs bit $j$ of all $r$ entries into $\ceil{r/w}$ words.
Two such vectors are added with word-parallel ripple-carry addition,
for all $r$ entries at once. For one word position, write the operand
planes as $A_j,B_j$ and set $C_0=0$. The output and carry planes are
\[
 H_j=A_j\mathbin{\oplus}B_j\mathbin{\oplus}C_j,\qquad
 C_{j+1}=(A_j\mathbin{\wedge}B_j)
 \mathbin{\vee}(C_j\mathbin{\wedge}(A_j\mathbin{\oplus}B_j)).
\]
Each bit position implements the usual full adder independently.
An extra plane keeps the final carry when the result needs one more
bit. With this representation, adding two vectors with $k$-bit entries therefore costs
$O(k\ceil{r/w})$ word operations, not $O(r)$.
Figure~\ref{fig:bit-planes} shows an example of the representation.

\paragraph{Pieces.}
To answer a query $R$, we split $R\cap\beta$, for each block $\beta$,
into $\ceil{|R\cap\beta|/s}$ \emph{pieces} of at most $s$ coordinates
each. Every piece is a set $S$ with a stored vector $V_S$. The sum over
$R\cap G$ is the sum of the vectors of the pieces inside $G$.
The number of pieces is
\begin{equation}\label{eq:pieces}
 p=\sum_{\beta}\ceil{|R\cap\beta|/s}\le |R|/s+b_R,
\end{equation}
where $b_R$ is the number of blocks meeting $R$. The inequality follows
from $\ceil{a/s}\le a/s+1$.

\begin{example}
Let $r=3$, $m=8$, $L=4$, and $s=2$. The blocks are $\set{0,1,2,3}$
and $\set{4,5,6,7}$. Let the rows of $M$ be $\set{0,3}$,
$\set{0,2,5}$, and $\set{3,5}$, and let $R=\set{0,2,3,5}$, all in one
group. The pieces are $\set{0,2}$ and $\set{3}$ in the first block and
$\set{5}$ in the second, so $p=3$. Their stored vectors are
\[
 V_{\set{0,2}}=(1,2,0),\quad V_{\set{3}}=(1,0,1),\quad
 V_{\set{5}}=(0,1,1),
\]
with sum $(2,3,2)$, which is $(|R\cap\text{row}_i|)_{i=1,2,3}$.
The entries are counts. Replacing the sum by a Boolean \textsc{or}
would give $(1,1,1)$, which records only that the intersections are
nonempty.
\end{example}

\begin{lemma}[Sparse tables]\label{lem:sparse}
The sparse table, consisting of the vectors $V_S$ for all nonempty
sets $S$ of at most $s$ coordinates within a block, occupies
$O((m/L)N_s(L)k\ceil{r/w})$ words. With incidence lists of total size
$O(mB_{s-1}(L-1))$, a previously zero entry $M[i,x]$ can be set to one
in $O(kB_{s-1}(L-1))$ time, keeping the table correct.
For a set $R$ of coordinates supplied as a sorted list, all group sums
$\sum_{x\in R\cap G}M[:,x]$ can be computed in
\[
 O\left(|R|+
 \bigl(pk+g_R\,(k+\lgv(|R|+1))\lgv(|R|+1)\bigr)\ceil{r/w}\right)
\]
time, where $g_R$ is the number of groups meeting $R$ and $p$ is the
number of pieces in~\eqref{eq:pieces}. The matrix is initially zero,
and initialization costs the stated storage and incidence bounds.
\end{lemma}

\begin{proof}
\emph{Space.} There are $m/L$ blocks, each with $N_s(L)$ stored
vectors, and each vector has $k$ planes of $\ceil{r/w}$ words.

\emph{Updates.} Fix a coordinate $x$. The stored sets that contain $x$
consist of $x$ and at most $s-1$ other coordinates of its block. There
are exactly $B_{s-1}(L-1)$ of them. The incidence list of $x$ stores
the addresses of their vectors. Setting $M[i,x]$ to one increments
entry $i$ of each of these vectors. One such increment reads and
updates at most $k$ planes. No entry exceeds $|S|\le s$, so $k$ planes
suffice.

\emph{Initialization.} The matrix is initially zero, so every stored
vector is zero: the tables are allocated and cleared. The incidence
lists are built by enumerating, for each block, the nonempty sets of at
most $s$ coordinates and appending the address of each set to the lists
of its members, in time proportional to their total size.

\emph{Queries.} We cut the sorted list
at block boundaries and split each block's part into pieces of at most
$s$ consecutive elements. To locate the vector of a piece, we compute its
rank: if the piece has local positions $p_1<\cdots<p_j$ within its
block, its rank among sets of size $j$ is $\sum_{i=1}^j\binom{p_i}{i}$,
plus an offset for the sets of smaller size. With a precomputed table
of binomial coefficients, all ranks take $O(|R|)$ time in total. Ranks
and table addresses fit in a word, since the tables fit in polynomial
space.
For a group $G$ with $p_G$ pieces, we add their vectors along a balanced
binary tree. 
At level $i$ of the tree, the operands have at most $k+i$
planes and there are at most $\ceil{p_G/2^{i+1}}$ additions. An unpaired
operand is passed up without arithmetic. Summing over levels, the
geometric part gives $O(p_Gk\ceil{r/w})$. The ceilings contribute at
most one extra addition per level, there are $O(\lgv(p_G+1))$ levels,
and the extra addition at level $i$ costs $O((k+i)\ceil{r/w})$; together
these give $O((k+\lgv(p_G+1))\lgv(p_G+1)\ceil{r/w})$. Since
$p_G\le|R|$ and the pieces partition $R\cap G$, summing over the $g_R$
groups gives the bound and the correct sums.
\end{proof}

\paragraph{Choosing the block length.}
We now relate $L$ to the promotion size $D$. Let $m_0$ be the size of
the original coordinate universe, before padding, and choose
$L=s\ceil{m_0/D}$. With $g$ groups, pad to a multiple of $gL$, so
$m<m_0+gL$. A promotion has $|R|=D$, and the number of blocks meeting
$R$ is at most $m/L<m_0/L+g\le D/s+g$. With
$B_s(L)\le(eL/s)^s$, this gives
\begin{equation}\label{eq:sparse-choice}
 N_s(L)\le(e\ceil{m_0/D})^s,
 \qquad p\le |R|/s+D/s+g.
\end{equation}
For $|R|=D$ the number of pieces is $O(D/s)$ whenever $g=O(D/s)$.
The additions of a promotion then cost $O((D/s)k\ceil{r/w})$ word
operations. Adding the $D$ columns one at a time would cost at least
$D\ceil{r/w}$.
In our application $m_0/D$ is polylogarithmic. Then
$N_s(L)$ is polynomial even for $s=\Theta(\lgv n/\lgv\lgv n)$.

Sparse subset tables are a variant of the data structure of Blelloch,
Vassilevska, and Williams~\cite[Theorem~1]{blelloch}. They store a
Boolean matrix and split its columns into blocks of length $\ell$.
For every block and every vector with at most $\kappa$ nonzeros, they
precompute the Boolean product of the block with that vector. A query
vector is split, within each block, into pieces with at most $\kappa$
nonzeros, which saves a factor of about $\kappa$ when $\ell$ is large.
Their structure also supports replacing a column. Our tables use the
same blocks, sparse subsets, and pieces, with $L=\ell$ and $s=\kappa$.
The difference is the stored values: we need exact counts, that is,
integer sums of columns, rather than Boolean products. This requires
the bit-plane representation, word-parallel carries, and the
integer-width accounting above. In addition, our updates set single
entries, which incidence lists handle at cost $O(kB_{s-1}(L-1))$ each.

\section{Two-sided incremental counting}\label{sec:sets}

This section completes the set-pair structure of
Section~\ref{sec:sparse} and proves its main guarantee,
Theorem~\ref{thm:sets}. 
A report of $X[v]\setminus Y[u]$ should scan only the groups that
contain an output. To find these groups, the structure maintains, for
every group $G$ and every pair of lines, the snapshot intersection
count
\[
 C_G[v,u]=|X_0[v]\cap Y_0[u]\cap G|.
\]
The section has four steps.
\begin{enumerate}
\item \emph{What is stored.} Besides the lines themselves, we store the
 sparse tables of Section~\ref{sec:sparse} for both snapshot matrices.
 We also store the counts $C_G[v,u]$, split into two counter arrays
 whose sum is $C_G[v,u]$ (invariant~\eqref{eq:invariant}).
\item \emph{Promotions.} A promotion updates the counts of the promoted
 line with all $n$ opposite lines at once. It uses a sparse-table
 query whose cost is given by Lemma~\ref{lem:sparse}.
\item \emph{Reports.} The counts, corrected by the two short pending
 lists, identify exactly the groups that contain an output. Only those
 groups are scanned.
\item \emph{Costs and parameters.} We bound the cost of each operation
 (Lemma~\ref{lem:setcost}), choose the parameters, and obtain
 Theorem~\ref{thm:sets}. 
\end{enumerate}

\paragraph{Intuition: counting when both families grow.}
The difficulty is that both families change, so neither family's
snapshots can be treated as fixed. Consider an element
$x\in X[v]\cap Y[u]\cap G$. It belongs to $C_G[v,u]$ exactly when it is
in both snapshots, that is, after the later of the two promotions that
move $x$ into $X_0[v]$ and into $Y_0[u]$. That later promotion finds $x$
already in the opposite snapshot. So the rule is simple: \emph{each
promotion counts its promoted elements against the current opposite
snapshots}. The earlier promotion does not count $x$, and the later
one does, so $x$ is counted exactly once. No coordination between lines
and no common snapshot time are needed.

The second question is where to store the counts. A promotion of
$Y[u]$ produces, for each group, a vector of $n$ counts indexed by
$v$, one for each opposite line $X[v]$. To add this vector to the
stored counts with one word-parallel vector addition
(Section~\ref{sec:sparse}), the $n$ counters for this fixed $u$ must
be stored contiguously. Thinking of $C_G$ as a matrix with rows $v$
and columns $u$, these counters form column $u$ of $C_G$.
Symmetrically, a promotion of $X[v]$ needs row $v$ of $C_G$ stored
contiguously. A single array cannot be laid out both ways. We
therefore keep two arrays: $I_{\rm col}$ receives the contributions
of $Y$ promotions, stored by columns, and $I_{\rm row}$ receives the
contributions of $X$ promotions, stored by rows. Their sum is $C_G$.

We store the following:
\begin{itemize}
\item Every line $X[v]$, $Y[u]$ and every snapshot $X_0[v]$, $Y_0[u]$
 is stored as a packed set, so that membership tests
 take constant time. Every pending list is also stored as a list of
 its elements.
\item Two sparse tables: $T_X$ for the
 snapshot matrix of the $X$ family, whose rows are the sets $X_0[v]$,
 and $T_Y$ for the snapshot matrix of the $Y$ family. Both use the
 same blocks and the same $g$ groups.
\item For every group $G$ and every $v$, the count
 $c_X[G][v]=|X_0[v]\cap G|$.
\item Two families of counter vectors. For every group $G$ and every
 $u$, a vector $I_{\rm col}[G][u]$ of $n$ counters indexed by $v$.
 For every group $G$ and every $v$, a vector $I_{\rm row}[G][v]$ of $n$
 counters indexed by $u$. Counters never exceed $n$, so each vector is
 stored in $b=\ceil{\lgv(n+1)}$ bit planes, as in
 Section~\ref{sec:sparse}.
\end{itemize}
The counter arrays satisfy the invariant
\begin{equation}\label{eq:invariant}
 I_{\rm col}[G][u][v]+I_{\rm row}[G][v][u]
 =|X_0[v]\cap Y_0[u]\cap G|=C_G[v,u].
\end{equation}
Initially all lines are empty and all stored quantities are zero, so
the invariant holds.

To insert an element into a line, we first test whether it is already
present; if so, we do nothing. Otherwise we add it to the line's packed set
and its pending list. If the pending list now has $D$ elements,
we promote the line as described next. 
Operations are executed one at a time, and a promotion finishes before the next operation begins. 
Hence every pending list has fewer than $D$ elements between operations.

\paragraph{Promoting a $Y$ line.}
Next, we describe promotions.
Let $R=\Delta_Y(u)$, so $|R|=D$. Lemma~\ref{lem:sparse} needs $R$ in
sorted order. Since $R=Y[u]\setminus Y_0[u]$ and both sets are stored
packed, enumerating their packed difference lists $R$ in increasing
order in $O(\ceil{n/w}+D)$ time (Lemma~\ref{lem:packed}).
Here is how promotions work:
\begin{enumerate}
\item Query $T_X$ with $R$ (see Lemma~\ref{lem:sparse} for costs). For every group
 $G$ this gives the vector $W_G$ with entries
 $W_G[v]=|X_0[v]\cap R\cap G|$ for all $v$.
\item Add $W_G$ to $I_{\rm col}[G][u]$ for every group $G$.
\item Move $R$ into $Y_0[u]$. In $T_Y$, set the entries $(u,x)$ for
 $x\in R$ to one, using the update operation of
 Lemma~\ref{lem:sparse}. Empty the pending list.
\end{enumerate}
Before the promotion, $R\cap Y_0[u]=\varnothing$. So the right side
of~\eqref{eq:invariant} increases by exactly $|X_0[v]\cap R\cap G|=W_G[v]$
for every $v$, and the left side increases by the same amount. The
$X$ snapshots do not change, so the invariant is preserved for all
other pairs.

\paragraph{Promoting an $X$ line.}
This is symmetric. Let $R=\Delta_X(v)$. Query $T_Y$ with $R$ to obtain
$W'_G[u]=|Y_0[u]\cap R\cap G|$ for all $u$. Add $W'_G$ to
$I_{\rm row}[G][v]$, and increase $c_X[G][v]$ by $|R\cap G|$. Then move
$R$ into $X_0[v]$, update $T_X$, and empty the pending list. The same
argument preserves~\eqref{eq:invariant}.
Promotions of the two families may interleave arbitrarily. The argument
holds for every interleaving, because each promotion counts against the
opposite snapshots exactly as they are just before it.

\begin{example}
Let an element $x$ belong to both $X[v]$ and $Y[u]$. Suppose $X[v]$
is promoted first. At that time $Y_0[u]$ does not contain $x$, so this
promotion adds zero for $x$. When $Y[u]$ is promoted later, $x$ is in
$X_0[v]$, and the promotion adds one to $I_{\rm col}$. In the opposite
order, the $X$ promotion comes second and adds one to $I_{\rm row}$.
Either way the sum in~\eqref{eq:invariant} counts $x$ exactly once.
\end{example}

\paragraph{Reporting a difference.}
We want $S=X[v]\setminus Y[u]$. Every output lies either in the pending
list $\Delta_X(v)$ or in the snapshot $X_0[v]$:
\begin{equation}\label{eq:report-split}
 X[v]\setminus Y[u]
 = (\Delta_X(v)\setminus Y[u])\;\dot\cup\;
   (X_0[v]\setminus Y[u]).
\end{equation}
The first part is found by scanning the short pending list. For the
second part, the stored counts tell how many elements of $X_0[v]\cap G$
are missing from $Y_0[u]$. Some of these may already be in $Y[u]$,
namely in its pending list $\Delta_Y(u)$. These must be subtracted,
which takes a scan of the short list $\Delta_Y(u)$. The procedure is as
follows.
\begin{enumerate}
\item Scan $\Delta_X(v)$. Mark the group of every element that is not
 in $Y[u]$.
\item For every group $G$, compute
 \[
  N_G=c_X[G][v]-I_{\rm col}[G][u][v]-I_{\rm row}[G][v][u].
 \]
 By~\eqref{eq:invariant},
 $N_G=|X_0[v]\cap G|-|X_0[v]\cap Y_0[u]\cap G|=|(X_0[v]\setminus Y_0[u])\cap G|$.
\item Scan $\Delta_Y(u)$ once, testing membership in $X_0[v]$, to
 compute $h_G=|X_0[v]\cap\Delta_Y(u)\cap G|$ for every group $G$. These
 elements are in $X_0[v]$ and in $Y[u]$ but not in $Y_0[u]$, so
 \begin{equation}\label{eq:remaining}
  N_G-h_G=|(X_0[v]\setminus Y[u])\cap G|\ge0.
 \end{equation}
\item Call a group $G$ \emph{live} if it was marked in step~1 or
 $N_G>h_G$. For each live group, compute the packed difference of
 $X[v]$ and $Y[u]$ restricted to the words of $G$ and list its
 elements (Lemma~\ref{lem:packed}).
\end{enumerate}
By~\eqref{eq:report-split} and~\eqref{eq:remaining}, a group is live
if and only if it contains an element of $S$. Hence the report lists
exactly $S$: every element of $S$ lies in a live group, and the scan of
a live group lists the elements of $S$ in it. Moreover, every scan of a
live group reports at least one output. We call such scans
\emph{productive}. In Section~\ref{sec:cfl}, every reported element is
a new fact, so the cost of each productive scan can be charged to a new
fact. This is how empty or nearly empty reports avoid paying for full
scans.

\begin{example}
Take groups $G_0=\set{0,1}$ and $G_1=\set{2,3}$, and lines with
$X_0[v]=\set{0,1}$, $\Delta_X(v)=\set{2}$,
$Y_0[u]=\set{0}$, and $\Delta_Y(u)=\set{1}$. Then $S=\set{2}$.
For $G_0$ the snapshot count is $N_{G_0}=1$, because $1\notin Y_0[u]$.
But $1\in\Delta_Y(u)$, so $h_{G_0}=1$ and $N_{G_0}-h_{G_0}=0$: the group
is not live and is not scanned. The pending element $2$ of $X[v]$ makes
$G_1$ live, and its scan returns exactly $2$. Without the correction
$h_G$, the group $G_0$ would be scanned without reporting anything, and
that scan could not be charged to an output.
\end{example}

\begin{lemma}[General costs]\label{lem:setcost}
A report operation on the data structure costs
\[
 O\left(D+g\lgv n+\sigma\ceil{n/(gw)}+|S|\right),
\]
where $\sigma\le\min(|S|,g)$ is the number of live groups.
If $g=O(D/s)$, a promotion costs
\[
 O\left(D+(D/s)k\ceil{n/w}
             +g\lgv^2 n\ceil{n/w}\right),
\]
excluding the sparse-table updates, which cost an additional
$O(kB_{s-1}(L-1))$ for every promoted element.
\end{lemma}
\begin{proof}
\emph{Reports.} The scans of the two pending lists cost $O(D)$.
Reading one counter from a vector stored in $b$ bit planes takes
$O(b)=O(\lgv n)$ time, and step~2 reads a constant number of counters
for each of the $g$ groups. Each group has $n/g$ coordinates after
padding, so scanning a live group costs $O(\ceil{n/(gw)})$ plus its
outputs (Lemma~\ref{lem:packed}).

\emph{Promotions.} Listing $R$ in sorted order costs
$O(\ceil{n/w}+D)$; this cost is absorbed by the other terms. 
We apply Lemma~\ref{lem:sparse} with $r=n$ and
$|R|=D$, and bound the number of pieces by~\eqref{eq:sparse-choice}.
With $g=O(D/s)$ this gives $p=O(D/s)$, which gives us the second and
third terms. Adding the $g$ resulting vectors into the counter arrays
costs $O(gb\ceil{n/w})$, which the last term absorbs. The table updates
are the update operation of Lemma~\ref{lem:sparse}, applied once per
promoted element.
\end{proof}

\paragraph{Parameter choice.}
Now we pick the  parameters to balance the costs in Lemma~\ref{lem:setcost}. The
promotion size $D$ must be large, because a promotion costs about
$(D/s)k\,n/w$ and is paid for by its $D$ insertions. This gives about
$nk/(sw)$ per insertion, independently of~$D$, but only if the other
promotion terms are smaller than this. $D$ must also be small, because
every report scans pending lists of size up to $D$. The choice
$D=\Theta(n/(\lgv n)^2)$ makes these scans as cheap as scanning one
group. With $g=\Theta(\lgv n)$ groups, reading the counters costs
$O((\lgv n)^2)$ and scanning one group costs $O(n/(\lgv n)^2)$. The
sparsity $s$ is as large as possible while each block has at most
$n^\eta$ stored vectors, which gives $s=\Theta(\lgv n/\lgv\lgv n)$. The
block length then follows from Section~\ref{sec:sparse}.

Precisely, let $\lambda=\ceil{\lgv n}$ and fix $\eta=\eps/4$. For
sufficiently large $n$, choose
\begin{equation}\label{eq:params}
 D=\ceil{n/\lambda^2},\quad g=\ceil{\lambda},\quad
 s=\floor{\frac{\eta\lgv n}{\lgv(e\ceil{n/D})}},\quad
 L=s\ceil{n/D}.
\end{equation}
Pad the universe to a multiple of $gL$, and add empty lines so that
there are as many lines as coordinates. The new size is
$n+O(gL)=\Theta(n)$, so all asymptotic bounds are unchanged. We have
\[
 s=\Theta(\lambda/\lgv\lambda),\quad k=\Theta(\lgv\lambda),
 \quad N_s(L)\le n^\eta,
 \quad gL=\polylog n=o(D).
\]
The last property implies $g=o(D/s)$, as Lemma~\ref{lem:setcost}
requires.

\begin{theorem}[Incremental set-pair structure]\label{thm:sets}
With the parameters above and $O(n^{2+\eps})$ initialization and space,
the following bounds hold.
\begin{itemize}
\item Inserting an element that is not yet in the line costs amortized
 $O(n(\lgv\lgv n)^2/(\eps(\lgv n)^2))$ time, for either family.
\item A report of $S=X[v]\setminus Y[u]$ costs
 $O((1+\sigma)n/(\lgv n)^2+|S|)$, where $\sigma\le\min(|S|,\Theta(\lgv n))$
 is the number of live groups.
\item The exact intersection count $|X[v]\cap Y[u]|$, and a test of
 whether $X[v]\subseteq Y[u]$, cost $O(n/(\lgv n)^2)$ in the worst case.
\end{itemize}
\end{theorem}
\begin{proof}
\emph{Space.} The two sparse tables occupy
$O((n/L)n^\eta k\ceil{n/w})$ words each, and their incidence lists
$O(n n^\eta)$. The two counter families occupy $O(gn^2 b/w)$ words.
Packed lines, lists, and group counts add at most $O(n^2+gn)$ words.
All polylogarithmic factors are absorbed by the slack $\eta<\eps$,
giving $O(n^{2+\eps})$ space and initialization.

\emph{Insertions.} Charge each promotion to the $D$ distinct
insertions in its pending list. By Lemma~\ref{lem:setcost}, the
dominant term per insertion is $O(nk/(sw))$. By~\eqref{eq:params},
$1/s=\Theta(\lgv\lambda/(\eta\lambda))$, $k=\Theta(\lgv\lambda)$, and
$w=\Theta(\lambda)$, so this term is
$O(n(\lgv\lambda)^2/(\eta\lambda^2))=O(n(\lgv\lambda)^2/(\eps\lambda^2))$.
The other terms per
insertion are $O(1+g\lgv^2 n\ceil{n/w}/D)$, which is
polylogarithmic. The table updates cost at most $O(kn^\eta)$ per
inserted element. Since $\eta<1$, all these terms are
$O(n(\lgv\lambda)^2/\lambda^2)$ for large $n$.

\emph{Reports.} This is Lemma~\ref{lem:setcost}, since
$g=\Theta(\lambda)$, $w=\Theta(\lambda)$, and $gb=O(\lambda^2)=O(D)$.

\emph{Intersection counts.} Use the disjoint union
\[
 X[v]\cap Y[u]
 =(X_0[v]\cap Y_0[u])\;\dot\cup\;
   (\Delta_X(v)\cap Y[u])\;\dot\cup\;
   (X_0[v]\cap\Delta_Y(u)).
\]
The size of the first part is the sum of~\eqref{eq:invariant} over all
groups. The other two parts are found by scanning the pending lists.
The total cost is $O(D+gb)=O(n/\lambda^2)$. Finally,
$X[v]\subseteq Y[u]$ holds if and only if $|X[v]\cap Y[u]|=|X[v]|$.
\end{proof}

\begin{remark}
If the $X$ lines are the rows of a Boolean matrix $A$ and the $Y$ lines
are the columns of a Boolean matrix $B$, both changing only from $0$
to $1$, then the structure maintains the integer product $AB$: each entry
$(AB)[v,u]=|X[v]\cap Y[u]|$ is available at the intersection-count
cost of Theorem~\ref{thm:sets}.
\end{remark}

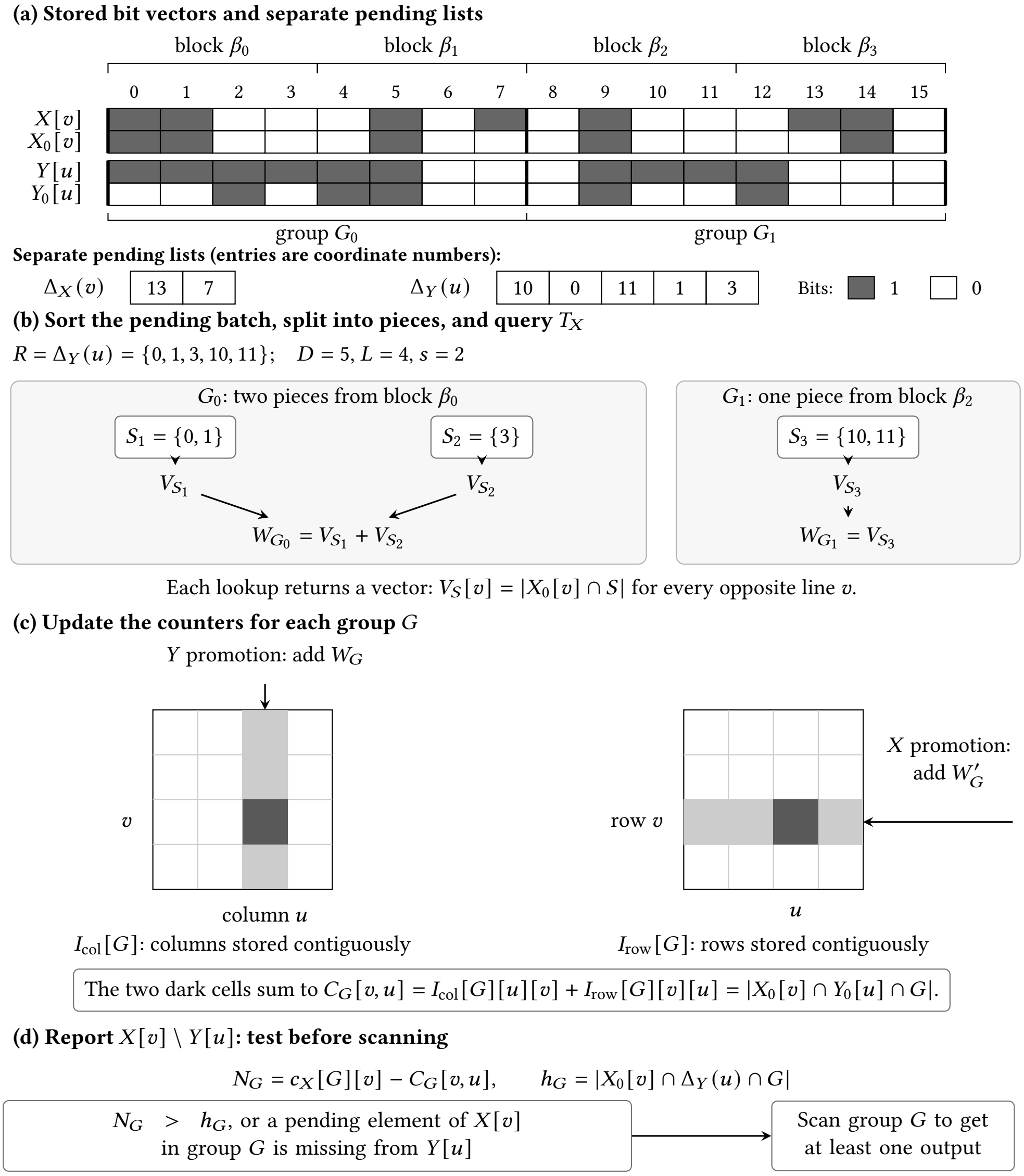
\begin{figure}[htbp]
\centering
\begin{tikzpicture}[
  x=0.95cm,y=-0.95cm,font=\footnotesize,>=stealth,
  title/.style={anchor=west,font=\footnotesize\bfseries},
  box/.style={draw=black!55,rounded corners=2pt,fill=white,inner sep=4pt},
  flow/.style={->,semithick}]
\node[title] at (0,0) {(a) Stored bit vectors and separate pending lists};
\foreach \b in {0,...,3} {
  \draw (1.4+2.8*\b,0.78) -- (1.4+2.8*\b,0.65)
    -- (4.2+2.8*\b,0.65) -- (4.2+2.8*\b,0.78);
  \node at (2.8+2.8*\b,0.43) {block $\beta_\b$};
}
\foreach \i in {0,...,15}
  \node[font=\scriptsize] at (1.75+0.7*\i,1.03) {$\i$};
\foreach \y/\members/\lab in {
  1.25/{0,1,5,7,9,13,14}/{X[v]},
  1.55/{0,1,5,9,14}/{X_0[v]},
  1.95/{0,1,2,3,4,5,9,10,11,12}/{Y[u]},
  2.25/{2,4,5,9,12}/{Y_0[u]}} {
  \foreach \i in \members
    \fill[black!60] (1.4+0.7*\i,\y) rectangle (2.1+0.7*\i,\y+0.3);
  \foreach \i in {0,...,15}
    \draw (1.4+0.7*\i,\y) rectangle (2.1+0.7*\i,\y+0.3);
  \node[anchor=east] at (1.2,\y+0.15) {$\lab$};
}
\foreach \i in {0,8,16} {
  \draw[very thick] (1.4+0.7*\i,1.25) -- (1.4+0.7*\i,1.85);
  \draw[very thick] (1.4+0.7*\i,1.95) -- (1.4+0.7*\i,2.55);
}
\foreach \g in {0,1} {
  \draw (1.4+5.6*\g,2.65) -- (1.4+5.6*\g,2.75)
    -- (7+5.6*\g,2.75) -- (7+5.6*\g,2.65);
  \node at (4.2+5.6*\g,2.94) {group $G_\g$};
}
\node[anchor=west,font=\scriptsize\bfseries] at (0,3.23)
  {Separate pending lists (entries are coordinate numbers):};
\node[anchor=east] at (1.5,3.65) {$\Delta_X(v)$};
\foreach \i/\value in {0/13,1/7} {
  \draw (1.7+0.7*\i,3.43) rectangle (2.4+0.7*\i,3.87);
  \node at (2.05+0.7*\i,3.65) {$\value$};
}
\node[anchor=east] at (6.4,3.65) {$\Delta_Y(u)$};
\foreach \i/\value in {0/10,1/0,2/11,3/1,4/3} {
  \draw (6.6+0.7*\i,3.43) rectangle (7.3+0.7*\i,3.87);
  \node at (6.95+0.7*\i,3.65) {$\value$};
}
\node[anchor=west,font=\scriptsize] at (10.5,3.65) {Bits:};
\filldraw[fill=black!60] (11.3,3.5) rectangle (11.65,3.8);
\node[anchor=west] at (11.72,3.65) {$1$};
\draw (12.4,3.5) rectangle (12.75,3.8);
\node[anchor=west] at (12.82,3.65) {$0$};

\node[title] at (0,4.1) {(b) Sort the pending batch, split into pieces, and query $T_X$};
\node[anchor=west] at (0,4.55)
  {$R=\Delta_Y(u)=\{0,1,3,10,11\}$;\quad $D=5$, $L=4$, $s=2$};
\draw[rounded corners=3pt,black!30,fill=black!3] (0.1,4.9) rectangle (8.6,7.35);
\draw[rounded corners=3pt,black!30,fill=black!3] (9,4.9) rectangle (13.6,7.35);
\node at (4.35,5.12) {$G_0$: two pieces from block $\beta_0$};
\node at (11.3,5.12) {$G_1$: one piece from block $\beta_2$};
\node[box] (s1) at (2.3,5.65) {$S_1=\{0,1\}$};
\node[box] (s2) at (6.4,5.65) {$S_2=\{3\}$};
\node[box] (s3) at (11.3,5.65) {$S_3=\{10,11\}$};
\node (v1) at (2.3,6.3) {$V_{S_1}$};
\node (v2) at (6.4,6.3) {$V_{S_2}$};
\node (v3) at (11.3,6.3) {$V_{S_3}$};
\draw[flow] (s1) -- (v1);
\draw[flow] (s2) -- (v2);
\draw[flow] (s3) -- (v3);
\node (w0) at (4.35,7) {$W_{G_0}=V_{S_1}+V_{S_2}$};
\node (w1) at (11.3,7) {$W_{G_1}=V_{S_3}$};
\draw[flow] (v1) -- (w0);
\draw[flow] (v2) -- (w0);
\draw[flow] (v3) -- (w1);
\node at (6.8,7.68)
  {Each lookup returns a vector: $V_S[v]=|X_0[v]\cap S|$ for every opposite line $v$.};

\node[title] at (0,8.15) {(c) Update the counters for each group $G$};
\node[anchor=south] at (3.5,8.85) {$Y$ promotion: add $W_G$};
\draw[flow] (3.5,8.95) -- (3.5,9.3);
\foreach \base in {2,9.1} {
  \draw (\base,9.3) rectangle (\base+2.4,11.7);
  \foreach \i in {1,2,3} {
    \draw[black!20] (\base+0.6*\i,9.3) -- (\base+0.6*\i,11.7);
    \draw[black!20] (\base,9.3+0.6*\i) -- (\base+2.4,9.3+0.6*\i);
  }
}
\fill[black!20] (3.2,9.3) rectangle (3.8,11.7);
\fill[black!65] (3.2,10.5) rectangle (3.8,11.1);
\fill[black!20] (9.1,10.5) rectangle (11.5,11.1);
\fill[black!65] (10.3,10.5) rectangle (10.9,11.1);
\node[anchor=east] at (1.85,10.8) {$v$};
\node[anchor=north] at (3.5,11.8) {column $u$};
\node[anchor=east] at (8.95,10.8) {row $v$};
\node[anchor=north] at (10.6,11.8) {$u$};
\node[align=center] at (12.65,10) {$X$ promotion:\\ add $W'_G$};
\draw[flow] (13.5,10.8) -- (11.5,10.8);
\node at (3.2,12.45) {$I_{\rm col}[G]$: columns stored contiguously};
\node at (10.3,12.45) {$I_{\rm row}[G]$: rows stored contiguously};
\node[box] at (6.8,13.05)
  {The two dark cells sum to $C_G[v,u]=I_{\rm col}[G][u][v]+I_{\rm row}[G][v][u]=|X_0[v]\cap Y_0[u]\cap G|$.};

\node[title] at (0,13.7) {(d) Report $X[v]\setminus Y[u]$: test before scanning};
\node at (6.8,14.25)
  {$N_G=c_X[G][v]-C_G[v,u]$,\qquad $h_G=|X_0[v]\cap\Delta_Y(u)\cap G|$};
\node[box,text width=7.7cm,align=center] (test) at (4.2,15)
  {$N_G>h_G$, or a pending element of $X[v]$\\ in group $G$ is missing from $Y[u]$};
\node[box,text width=2.8cm,align=center] (scan) at (11.9,15)
  {Scan group $G$ to get\\ at least one output};
\draw[flow] (test) -- (scan);
\end{tikzpicture}
\caption{The set-pair structure. (a)~Current sets and snapshots are
stored as separate packed bit vectors. Pending members are already set
in the current vector, absent from its snapshot, and also recorded in
an explicit list below. The fifth pending $Y$ insertion triggers a
promotion; other pending lists have fewer than $D$ elements. (b)~The
pending coordinates are enumerated in sorted order and split into
pieces of at most $s=2$ coordinates within a block. Sparse-table vectors are added
separately for each group; for the displayed $X[v]$ line,
$W_{G_0}[v]=2$ and $W_{G_1}[v]=0$. (c)~A $Y$ promotion adds a column vector;
an $X$ promotion symmetrically uses $T_Y$ and adds a row vector.
The matrices show the two storage orientations schematically, and the
dark cells represent the same pair $(v,u)$. After a promotion, its
pending elements enter its snapshot and its own sparse table is updated.
(d)~A report reads the two counters and corrects for pending insertions
before scanning a group. A group that fails the test is skipped.}
\Description{Four panels show storage, promotion, counter updates,
and reporting. Four separate bit vectors represent the current X and Y
sets and their snapshots; filled cells are one bits and empty cells are
zero bits. Below them, explicit pending lists contain X coordinates
13 and 7, and Y coordinates 10, 0, 11, 1, and 3. These members are also
set in the current bit vectors but not in the snapshots. The Y batch
is sorted and split into pieces 0 and 1, then 3, then 10 and 11.
Arrows connect these pieces to table vectors and per-group sums.
Two matrices highlight a column and a row, with a dark cell in each
whose values sum to the snapshot intersection count. A final test
corrects this count using pending entries and scans only groups that
contain an output.}
\label{fig:schematic}
\end{figure}

Figure~\ref{fig:schematic} shows one set-pair instance. The three
ingredients are the two families of lines with their snapshots and
pending lists (Section~\ref{sec:sparse}), the sparse tables that turn a
sorted pending list into per-group count vectors with $O(D/s)$ table
lookups instead of $D$ column additions (Lemma~\ref{lem:sparse}), and
the two counter arrays whose opposite layouts let both kinds of
promotion add their vectors contiguously.
The report never reads the sparse tables. 
It reads $O(g)$ counters, scans
the two pending lists, and touches the packed words of a group only when
the corrected count or a pending element guarantees an output there.

\section{The all-pairs CFL reachability algorithm}\label{sec:cfl}

This section combines \textsc{Saturate} (Section~\ref{sec:prelim})
with the set-pair structure (Theorem~\ref{thm:sets}) and proves the
main theorem. It has three steps.
\begin{enumerate}
\item \emph{Instantiation.} Each binary production gets two set-pair
 instances, one for the row differences and one for the column
 differences of \textsc{Saturate}. Every insertion into a relation is
 forwarded to the instances that contain it.
\item \emph{Charging.} Each report costs $O(n/(\lgv n)^2)$, plus
 $O(n/(\lgv n)^2)$ for each live group, plus its output. There can be
 $\Theta(n^2)$ reports with $\Theta(\lgv n)$ live groups each, so
 bounding each report separately is too weak. Instead, every live group
 contains a new fact, so the total number of live groups over the
 whole run is at most the number of facts.
\item \emph{Running time.} Summing the costs of Theorem~\ref{thm:sets}
 with this charging gives the main theorem.
\end{enumerate}

\paragraph{Set-pairs.}
For every binary production $A\to BC$, we create two instances of the
set-pair structure:
\begin{center}
\begin{tabular}{lllll}
\toprule
Instance & $X$ lines & $Y$ lines & Processed fact & Report\\
\midrule
Row instance & rows of $R_C$ & rows of $R_A$ & $B(u,v)$ &
 $R_C[v,:]\setminus R_A[u,:]$\\
Column instance & columns of $R_B$ & columns of $R_A$ & $C(v,x)$ &
 $R_B[:,v]\setminus R_A[:,x]$\\
\bottomrule
\end{tabular}
\end{center}
The row instance reports the row difference~\eqref{eq:diff-trigger},
and the column instance reports the column difference. Each instance
keeps its own copies of its lines. A relation $R_Z$ can occur in
several instances, by rows or by columns and as the $X$ or the $Y$
family. Relations only grow during saturation. This matches the
set-pair structure, which supports insertions but no deletions.

\paragraph{Instantiating saturation.}
We run \textsc{Saturate} (Algorithm~\ref{algo:saturate}) with two
changes.
\begin{itemize}
\item After \textsc{Insert}$(Z,a,b)$ sets the bit $(a,b)$ of $R_Z$, it
 \emph{notifies} every instance in which $R_Z$ is a family. If the
 instance uses the rows of $R_Z$, the notification inserts $b$ into line
 $a$. If it uses the columns, it inserts $a$ into line $b$. Such an
 insertion may cause a promotion inside the instance. A promotion
 changes only the instance's internal data. It does not call
 \textsc{Insert} and does not touch the worklist.
\item Each difference $D$ in \textsc{Saturate} is computed by a report
 query to the corresponding instance.
\end{itemize}
The full pseudocode is given in the supplementary material.
As in \textsc{Saturate}, a report is completed before any of its
elements are inserted. This matters here because the insertions
notify the same instance, and they would otherwise change the $Y$ line
being subtracted, or trigger a promotion, during the report.

Every insertion into a relation is forwarded to all instances
containing it. So, at all times, the lines of every instance are the
current rows or columns of the corresponding relations. A report
returns exactly the requested difference (Section~\ref{sec:sets}).
Hence the run computes the same differences as \textsc{Saturate}, and
Lemma~\ref{lem:saturate} applies.

\begin{lemma}[All-pairs correctness]\label{lem:cflcorrect}
The algorithm terminates and, for every nonterminal $A$ and every
ordered pair $(u,v)\in V\times V$, its returned matrix satisfies
$R_A[u,v]=1$ if and only if some walk from $u$ to $v$ has a label in
$L_A$. In particular, the returned $(R_A)_{A\in N}$ solves all-pairs CFL
reachability.
\end{lemma}

\paragraph{Charging.}
Let $F=\sum_{A\in N}|R_A|$ be the number of facts in the final
relations, including the initial facts. This counts facts over all ordered vertex
pairs, not only those relevant to a designated query. For a fixed
grammar, we have $F=O(n^2)$. Fix the row instance of a production $A\to BC$;
the column instance is symmetric.
\begin{itemize}
\item \emph{Insertions.} Its lines receive one insertion for each fact
 of $R_C$ and one for each fact of $R_A$, so at most $2F$ insertions.
 (The two relations may coincide, when $C=A$.)
\item \emph{Reports.} There is one report for each processed fact
 $B(u,v)$, so at most $F$ reports.
\item \emph{Promotions.} Each promotion empties a pending list of $D$
 elements, so there are at most $2F/D$ promotions.
\item \emph{Live groups.} Each live group of a report contains at least
 one reported element $x$ (Section~\ref{sec:sets}). Such an element
 is a new fact $A(u,x)$: it is absent from $R_A$ when the report starts
 and is inserted right after the report. Afterwards $x$ is in the $Y$
 line $R_A[u,:]$, so no later report of this instance returns it again.
 Hence the total number of live groups, summed over all reports of the
 instance, is at most $|R_A|\le F$.
\end{itemize}
The last item is the key to the analysis. Bounding each report by its
worst case of $g$ live groups would give a total of $Fg$ group scans,
which is too large.

\begin{theorem}[Refined all-pairs saturation bound]\label{thm:refined}
For a Chomsky normal form grammar with $P$ binary productions, size
$|\mathcal G|$, and $F$ final nonterminal facts, saturation computes all
nonterminal relations $R_A$ and takes
\[
 O\left(P\left(n^{2+\eps}
 +\frac{Fn(\lgv\lgv n)^2}{\eps(\lgv n)^2}\right)
 +|\mathcal G|n^2\right)
\]
time and $O(Pn^{2+\eps}+|\mathcal G|n^2)$ words of space. The
constants in the $O$ do not depend on $\eps$.
\end{theorem}
\begin{proof}
Fix one instance. By Lemma~\ref{lem:setcost} and the counts above,
its cost besides initialization is
\begin{align*}
 O\bigg(&F(D+g\lgv n)+F\ceil{n/(gw)}+F\\
 &+\frac FD\left(D+\frac Ds k\ceil{n/w}
                      +g\lgv^2 n\ceil{n/w}\right)
 +Fk n^\eta\bigg).
\end{align*}
The first line covers the reports: the fixed part of each of the at
most $F$ reports, the scans of at most $F$ live groups in total, and
at most $F$ reported elements. The second line covers the $O(F/D)$
promotions and the table updates. Each inserted element is
promoted once, and its table updates cost
$O(kB_{s-1}(L-1))=O(kn^\eta)$.
With~\eqref{eq:params}, the leading term is the promotion term
$O(Fnk/(sw))=O(Fn(\lgv\lambda)^2/(\eta\lambda^2))$ with
$\eta=\eps/4$ (see the proof of Theorem~\ref{thm:sets}). The
remaining terms are, per fact,
$O(n/\lambda^2+\polylog n+kn^\eta)$, all dominated by
$O(n(\lgv\lambda)^2/\lambda^2)$.

The two instances of each binary production are initialized in
$O(n^{2+\eps})$ time, for $O(Pn^{2+\eps})$ in total. Seeding terminal
and nullable facts and processing unit rules cost $O(|\mathcal G|n^2)$,
apart from the notifications already counted: each unit rule $A\to B$
is applied once to each of the at most $n^2$ facts of $R_B$, including
attempts that find the fact already present. Summing over all
productions proves the time bound. The space bound adds the
$O(n^{2+\eps})$ space of each instance (Theorem~\ref{thm:sets}) to the
$O(|\mathcal G|n^2)$ space of the relations and the worklist.
\end{proof}

\begin{proof}[Proof of Theorem~\ref{thm:main-intro}]
Lemma~\ref{lem:cflcorrect} establishes correctness for every entry of
all the returned matrices. Apply Theorem~\ref{thm:refined} with a
fixed grammar and $F=O(n^2)$ to bound their computation. Returning
$R_S$, or all the nonterminal matrices, requires no further saturation;
even copying their packed representation costs only $O(n^2)$ time.
Since $\eps<1$, initialization and all lower-order polynomial and
polylogarithmic terms are eventually dominated by the stated time.
\end{proof}

The two factors $\lgv\lgv n$ in the bound have different origins. One
is the inverse sparsity $1/s=\Theta(\lgv\lgv n/\lgv n)$. It arises
because $s=\eta\lgv n/\lgv(e\ceil{n/D})$ and $n/D=\Theta((\lgv n)^2)$
(equation~\eqref{eq:params}). The other is the number $k$ of bit
planes needed for an exact count of a piece. Improving either one
would improve the final bound.

The two counter families occupy only $O(gn^2b/w)=O(n^2\lgv n)$ words.
The exponent $\eps$ in the space accounting comes entirely from the two sparse tables, whose
size is $\Theta((n/L)N_s(L)k\ceil{n/w})$ words. 
A smaller sparsity trades time for space. 
With $s=\floor{\lambda/(\lgv\lambda)^2}$ in
place of the choice in~\eqref{eq:params}, the bound
$B_s(L)\le(e\ceil{n/D})^s$ gives
$N_s(L)\le2^{O(\lambda/\lgv\lambda)}=n^{O(1/\lgv\lgv n)}$, so the
structure uses $n^{2+o(1)}$ words, while the amortized insertion cost
$O(nk/(sw))$ becomes $O(n(\lgv\lgv n)^3/(\lgv n)^2)$. The proofs
above use only $N_s(L)\le n^\eta$ for some $\eta<1$ and
$gL=o(D)$, so they apply unchanged. Consequently, the algorithm of
Section~\ref{sec:cfl} also solves all-pairs CFL reachability for a
fixed grammar in $O(n^3(\lgv\lgv n)^3/(\lgv n)^2)$ time and
$n^{2+o(1)}$ space.

\paragraph{Relation to fast matrix--vector products.}
The connection with known fast matrix--vector
algorithms~\cite{williams-mv,larsen-williams,liu} can be seen from a
single worklist event. Recall that processing a fact $B(u,v)$ for
a rule $A\to BC$ asks for destinations in $X[v]\setminus Y[u]$:
those reachable through $C$ from $v$ whose corresponding $A$ fact from
$u$ is still missing. A Boolean matrix--vector product can test whether
this difference is nonempty. Take the matrix whose row $u$ marks the
elements missing from $Y[u]$, and multiply it by the characteristic
vector of $X[v]$. The product answers the nonemptiness question for
every $u$ at once, although this worklist event asks about only one
$u$. There are $O(n^2)$ worklist queries, so invoking a whole product
for each query does not give a better bound. 
Even dynamic algorithms that support changing matrices and adaptive
queries~\cite{liu} need a way to share this work across many events.

Batching updates offers such sharing, but then the stored answers
describe snapshots, and queries must account for insertions that have
not yet been included. The difficulty already appears when $X[v]$
stays fixed and only $Y[u]$ grows. Suppose the snapshot difference is
either $\set{x}$ or $\set{x,y}$. A Boolean answer records ``nonempty''
in both cases. If $x$ is then inserted into $Y[u]$, the first
difference becomes empty, while the second still contains $y$. The
stored Boolean answer cannot distinguish these outcomes, and storing
$x$ as a witness does not help. A stored count does: it starts at one
or two, respectively, and subtracting the contribution of the pending
insertion gives zero or one. Our structure performs this correction
within each group by inspecting the short pending lists.

Maintaining exact counts is itself expensive if every insertion
updates its counts against all opposite lines separately. Our sparse
tables make these updates affordable by processing a batch of new
elements together. The corrected counts then identify precisely which
groups contain a missing fact. Scanning each such group discovers at
least one new fact, so the $O(n^2)$ bound on new facts also bounds the
total number of these scans. Thus the improvement combines efficient
batch updates, correction of snapshot answers, and a bound on the work
spent discovering new facts. 
We do not know if existing matrix-vector algorithms can be modified to provide
these combinations.

\section{2NPDA recognition}\label{sec:machines}

A \emph{2NPDA} has finite state and stack alphabets, reads a fixed input
tape in either direction, uses an unbounded stack, and has a
nondeterministic transition relation \cite{HarrisonI68}.
Recall that the recognition problem for a \emph{fixed} 2NPDA asks whether
it accepts a given input word of length $\ell$.
Chistikov, Majumdar, and Schepper show linear-time reductions between 2NPDA recognition and 
designated-pair Dyck-2 reachability in the bit-size of the input~\cite[Proposition 20 and Appendix E]{cms}.
Since an explicitly encoded dense graph may require
$\Theta(n^2\lgv n)$ bits, these equivalences cannot equate the
length of an input word with graph vertex count. 
Therefore, we give a direct $O(\ell)$-vertex construction needed for our recognition bound.
Our all-pairs algorithm yields recognition by answering the designated queries arising in that construction.

Fix a 2NPDA $M$ with states $Q$, stack alphabet $\Gamma$, and
bottom symbol $Z_0$. Its input tape contains $\vdash x\dashv$ with positions
$0,\ldots,\ell+1$, where $\vdash$ and $\dashv$ are special end-of-tape markers. 
A transition tests the current state, tape symbol,
and stack top, changes the state, moves by $-1$, $0$, or $1$, and performs a push or pop on the stack or leaves the stack as it is.
The head stays within the endmarkers and $Z_0$ cannot be popped. 
Any 2NPDA can be normalized this way. For conventional final-state acceptance,
reserve a bottom symbol that is never pushed and append a clearing
phase. We use acceptance in a final state with only the bottom remaining.

We now reduce 2NPDA recognition to (designated-pair) CFL reachability.
The graph has vertices $(q,i,Z)\in Q\times[\ell+2]\times\Gamma$.
For a legal transition at position $i$, ending in state $q'$ and position
$i+d$, $d\in \{-1, 0, 1\}$, we introduce the following edges:
\begin{center}
\begin{tabular}{ll}
\toprule
No change & $(q,i,Z)\xrightarrow{e}(q',i+d,Z)$\\
Push $Z'$ & $(q,i,Z)\xrightarrow{o_Z}(q',i+d,Z')$\\
Pop & $(q,i,Z)\xrightarrow{c_Y}(q',i+d,Y)$ for every $Y\in\Gamma$\\
\bottomrule
\end{tabular}
\end{center}
A pop edge guesses the revealed symbol. 
We use a fixed grammar, a Dyck grammar with one parenthesis type per
stack symbol and a neutral letter $e$, to check
this guess by recording the old top through the opening label:
$S\to\eps\mid e\mid SS\mid o_YSc_Y$ for $Y\in\Gamma$.

\begin{lemma}[Flat computations]\label{lem:flat}
There is an $S$-labeled walk from $(q,i,Z)$ to $(q',i',Z')$ if and only
if $Z=Z'$ and there is a computation from stack $\alpha Z$ to that
same stack, with these control states and head positions, that never
pops the displayed $Z$. 
\end{lemma}
\begin{proof}
Transitions inspect only the top, so a computation preserving the
displayed prefix is independent of $\alpha$. For the forward direction,
induct on the grammar derivation. Empty and $e$ walks give the empty
computation and a no-change step. Concatenation composes flat computations.
For $o_YSc_Y$, the initial surface top must be $Y$. The opening edge
pushes some $Z''$, and the inductive interior begins and ends with top
$Z''$ without removing it. The final pop exposes $Y$ and reaches the
specified surface. Thus the guessed pop destination is correct.

Conversely, induct on the length of a flat computation. The first step
cannot pop its protected top. A no-change step gives an $e$ edge and
a shorter flat remainder. After an initial push of $Z''$, take the first
step returning to the original height. Before that pop, the computation
is flat at $Z''$; apply induction to this interior and to the remainder.
The outer edges are $o_Z$ and $c_Z$, yielding $o_ZSc_Z$ followed by
the remainder. This proves the grammar derivation.
\end{proof}

\begin{corollary}\label{cor:2npda}
Recognition by any fixed 2NPDA can be solved in
$O(\ell^3(\lgv\lgv\ell)^2/(\lgv\ell)^2)$ time and
$O(\ell^{2+\eps})$ words of space.
\end{corollary}
\begin{proof}
The graph has $|Q||\Gamma|(\ell+2)=O(\ell)$ vertices and $O(\ell)$
edges and is constructed in linear time. By Lemma~\ref{lem:flat},
acceptance is precisely an $S$-walk from $(q_0,0,Z_0)$ to some
$(q_f,i,Z_0)$ with $q_f$ final. Apply the all-pairs algorithm of
Theorem~\ref{thm:main-intro} to this graph and test these $O(\ell)$
entries of the returned relation. This recognition consequence uses
only designated entries of an output that is computed for all pairs.
\end{proof}

\section{Discussion and open problems}

We have made the first progress on the running time of context-free
reachability and 2NPDA recognition since Rytter's packed
sets~\cite{rytter,chaudhuri}: we show an asymptotic improvement by a 
$\lgv n/(\lgv\lgv n)^2$ factor.
The key innovations are to use a novel sparse-set data structure and to batch
updates to the data structure within a line.
A natural follow-up is to remove the $(\lgv\lgv n)^2$ factor to get a ``pure''
$O(n^3/(\lgv n)^2)$ algorithm.
A more ambitious goal will be to match the best algorithms for NFA acceptance.
While our improvements are theoretical, we hope that our ideas of using
sparse sets can also lead to more efficient implementations in practice.

\section*{AI Use}

We used Claude Opus 5.5 to understand the key ideas of \cite{blelloch,williams-mv,omv,larsen-williams} and why
they do not show an improvement for CFL reachability.
We used Claude Opus 5.5 to iterate over the design of the data structure and its proof of correctness.
We wrote our own exposition of the results starting from those proofs.
We used Claude Fable 5.1 to autoformalize the main results from this draft except for the word-RAM model.
The formalization found no errors.

\newpage

%

\bibliographystyle{ACM-Reference-Format}
\bibliography{references}

\clearpage

\appendix

\section{Pseudocode and a worked example}
\label{app:illustrations}

The appendix shows the pseudocode for the four operations on the set-pair structure
(Algorithm~\ref{algo:setpair}) and a small saturation run in which
the counts let one report finish without scanning any group.

\subsection{Pseudocode}

Algorithm~\ref{algo:setpair} gives the four operations of one set-pair
instance in the notation of Section~\ref{sec:sets}. We write
$\mathrm{grp}(x)$ for the group containing coordinate $x$, and
$V^X_S$ and $V^Y_S$ for the stored vectors of the tables $T_X$ and $T_Y$.
\textsc{Promote}$(X,v)$ is obtained from \textsc{Promote}$(Y,u)$ by
exchanging the roles of the two families, with the two differences
noted in its description. In Section~\ref{sec:cfl}, \textsc{Insert} of
Algorithm~\ref{algo:saturate} calls \textsc{Insert} of this algorithm
on every instance containing the updated relation, and each difference
of Algorithm~\ref{algo:saturate} is one call to \textsc{Report}.

\begin{algorithm}[t]
\begin{quote}\small
\textsc{Insert}$(F,\ell,x)$:\quad\emph{(family $F\in\set{X,Y}$, line $\ell$, element $x$)}\\
\hspace*{1em}If $x\in F[\ell]$, return.\\
\hspace*{1em}Add $x$ to the packed set $F[\ell]$ and to the pending list $\Delta_F(\ell)$.\\
\hspace*{1em}If $|\Delta_F(\ell)|=D$, call \textsc{Promote}$(F,\ell)$.\\[3pt]
\textsc{Promote}$(Y,u)$:\\
\hspace*{1em}$R\gets$ the elements of $Y[u]\setminus Y_0[u]$ in increasing order (packed difference).\\
\hspace*{1em}Cut $R$ at block boundaries; split each part into pieces of at most $s$ consecutive elements.\\
\hspace*{1em}For each group $G$: $W_G\gets\sum\set{V^X_S : S\text{ a piece inside }G}$,
 added along a balanced tree of bit-plane additions.\\
\hspace*{1em}For each group $G$: $I_{\rm col}[G][u]\gets I_{\rm col}[G][u]+W_G$
 \quad\emph{(one vector addition of $n$ counters)}.\\
\hspace*{1em}$Y_0[u]\gets Y_0[u]\cup R$; for each $x\in R$ set entry $(u,x)$ of $T_Y$ to one; $\Delta_Y(u)\gets\varnothing$.\\[3pt]
\textsc{Promote}$(X,v)$:\quad as above with $X\leftrightarrow Y$, $T_Y$ in place of $T_X$,
 $I_{\rm row}[G][v]\gets I_{\rm row}[G][v]+W'_G$,\\
\hspace*{1em}and additionally $c_X[G][v]\gets c_X[G][v]+|R\cap G|$ for each group $G$.\\[3pt]
\textsc{Report}$(v,u)$:\quad\emph{(returns $S=X[v]\setminus Y[u]$)}\\
\hspace*{1em}$\mathit{Live}\gets\varnothing$; $S\gets\varnothing$.\\
\hspace*{1em}For each $x\in\Delta_X(v)$: if $x\notin Y[u]$, add $\mathrm{grp}(x)$ to $\mathit{Live}$.\\
\hspace*{1em}For each group $G$: $N_G\gets c_X[G][v]-I_{\rm col}[G][u][v]-I_{\rm row}[G][v][u]$; $h_G\gets0$.\\
\hspace*{1em}For each $y\in\Delta_Y(u)$: if $y\in X_0[v]$, increment $h_{\mathrm{grp}(y)}$.\\
\hspace*{1em}For each group $G$: if $N_G>h_G$, add $G$ to $\mathit{Live}$.\\
\hspace*{1em}For each $G\in\mathit{Live}$: append to $S$ the elements of $(X[v]\setminus Y[u])\cap G$,
 found by a packed scan of the words of $G$.\\
\hspace*{1em}Return $S$.\\[3pt]
\textsc{Count}$(v,u)$:\quad\emph{(returns $|X[v]\cap Y[u]|$)}\\
\hspace*{1em}Return $\sum_G\bigl(I_{\rm col}[G][u][v]+I_{\rm row}[G][v][u]\bigr)
 +|\set{x\in\Delta_X(v):x\in Y[u]}|+|\set{y\in\Delta_Y(u):y\in X_0[v]}|$.
\end{quote}
\caption{The operations of one set-pair instance. \textsc{Report}
scans the packed words of a group only when that group is live, and
every live group contains an output. $X[v]\subseteq Y[u]$ holds if and
only if \textsc{Count}$(v,u)=|X[v]|$.
\label{algo:setpair}}
\end{algorithm}

\subsection{A worked saturation run}

We trace one row instance on a small input. The parameters are chosen
for readability, not from~\eqref{eq:params}: eight vertices
$V=\set{0,\dots,7}$, two groups $G_0=\set{0,1,2,3}$ and
$G_1=\set{4,5,6,7}$, blocks equal to groups ($L=4$), sparsity $s=2$,
and promotion threshold $D=2$. The grammar has the rules
$S\to AB$, $A\to a$, and $B\to b$. The graph has the edges
\[
 (0,a,1),\ (0,a,2),\qquad
 (1,b,3),\ (1,b,5),\ (1,b,6),\ (2,b,3),\ (2,b,6),
\]
listed in the order in which \textsc{Saturate} inserts them. We follow
the row instance of $S\to AB$: its $X$ lines are the rows of $R_B$,
its $Y$ lines are the rows of $R_S$, and processing a fact $A(u,v)$
asks for $R_B[v,:]\setminus R_S[u,:]$, that is, \textsc{Report}$(v,u)$.
The sparse table $T_X$ stores, for each block, the vectors $V_S$ of all
nonempty $S$ with $|S|\le2$ inside the block, ten vectors per block,
each with eight counters (one per $X$ line) in $k=2$ bit planes.

\paragraph{Seeding.}
The two $A$ facts enter the worklist and do not touch this instance.
The $B$ facts are forwarded as insertions into $X$ lines.
Line $X[1]$ receives $3$ and $5$; its pending list reaches $D=2$, so
\textsc{Promote}$(X,1)$ runs with $R=\set{3,5}$. The table $T_Y$ is
still all zero, so $W'_{G_0}=W'_{G_1}=0$ and $I_{\rm row}$ stays zero;
the promotion sets $X_0[1]=\set{3,5}$, $c_X[G_0][1]=c_X[G_1][1]=1$, and
records $3$ and $5$ in $T_X$. Then $X[1]$ receives $6$, which stays
pending: $\Delta_X(1)=\set{6}$. Line $X[2]$ receives $3$ and $6$ and is
promoted the same way: $X_0[2]=\set{3,6}$,
$c_X[G_0][2]=c_X[G_1][2]=1$, $\Delta_X(2)=\varnothing$. All $Y$ lines are
empty. After seeding, the table entries relevant below are
\[
 V_{\set3}=(0,1,1,0,0,0,0,0),\qquad
 V_{\set5}=(0,1,0,0,0,0,0,0),
\]
indexed by $v=0,\dots,7$: coordinate $3$ lies in $X_0[1]$ and $X_0[2]$,
coordinate $5$ only in $X_0[1]$.

\paragraph{Processing $A(0,1)$: a report with outputs.}
\textsc{Report}$(1,0)$ computes $X[1]\setminus Y[0]=\set{3,5,6}$.
The pending element $6\in\Delta_X(1)$ is not in $Y[0]$, so $G_1$ is
marked live. The counts are $N_{G_0}=c_X[G_0][1]-0-0=1$ and
$N_{G_1}=1$; the list $\Delta_Y(0)$ is empty, so $h_{G_0}=h_{G_1}=0$ and
both groups are live. The scan of $G_0$ returns $3$ and the scan of
$G_1$ returns $5$ and $6$. Both scans are productive.
\textsc{Saturate} now inserts $S(0,3)$, $S(0,5)$, and $S(0,6)$. The
first two insertions fill $\Delta_Y(0)$ to size $D=2$, which triggers
\textsc{Promote}$(Y,0)$ with $R=\set{3,5}$. The sorted list is cut at
the block boundary into the pieces $\set3$ and $\set5$, whose vectors
are read from $T_X$:
\[
 W_{G_0}=V_{\set3}=(0,1,1,0,\dots),\qquad
 W_{G_1}=V_{\set5}=(0,1,0,0,\dots).
\]
These are added to column $0$ of $I_{\rm col}[G_0]$ and
$I_{\rm col}[G_1]$, so now $I_{\rm col}[G_0][0][1]=I_{\rm col}[G_0][0][2]=1$
and $I_{\rm col}[G_1][0][1]=1$, all other counters being zero. The
promotion sets $Y_0[0]=\set{3,5}$. This is the two-sided rule at work:
the $X$ promotions came first and counted nothing, and the later $Y$
promotion counted each shared element once. Finally $S(0,6)$ is
inserted and stays pending, $\Delta_Y(0)=\set6$.

\paragraph{Processing $A(0,2)$: an empty report without a scan.}
\textsc{Report}$(2,0)$ asks for $X[2]\setminus Y[0]=\set{3,6}\setminus\set{3,5,6}$,
which is empty. The pending list $\Delta_X(2)$ is empty, so no group
is marked. The counts are
\[
 N_{G_0}=c_X[G_0][2]-I_{\rm col}[G_0][0][2]-I_{\rm row}[G_0][2][0]=1-1-0=0,
 \qquad
 N_{G_1}=1-0-0=1.
\]
The snapshot count for $G_1$ is positive because $6\in X_0[2]$ is not
in the snapshot $Y_0[0]$. The scan of $\Delta_Y(0)=\set6$ finds
$6\in X_0[2]$, so $h_{G_1}=1$ and $N_{G_1}-h_{G_1}=0$. No group is
live. The report returns $\varnothing$ after reading six counters and
two pending lists of total length one, without touching the packed
words of either line. Without the correction $h_{G_1}$, group $G_1$
would have been scanned for nothing, and that scan could not be charged
to a new fact. As a check, \textsc{Count}$(2,0)=(1+0)+(0+0)+0+1=2=|X[2]|$,
confirming $X[2]\subseteq Y[0]$.

\paragraph{The remaining facts.}
The five $B$ facts are processed next. For the rule $S\to AB$ they
query the column instance, whose reports
$R_A[:,v]\setminus R_S[:,x]$ are all empty because every $S$ fact was
already found above; they are handled exactly as the report just
described. The worklist then contains $S(0,3)$, $S(0,5)$, and $S(0,6)$,
which match no rule, and saturation ends with
$R_S=\set{(0,3),(0,5),(0,6)}$.

\end{document}